\documentclass[journal,draftclsnofoot,onecolumn,12pt]{IEEEtran}  
\usepackage{amsthm,amssymb,graphicx,multirow,amsmath,color,amsfonts}

\usepackage[update,prepend]{epstopdf}
\usepackage[noadjust]{cite}

\usepackage[latin1]{inputenc}
\usepackage{tikz}
\usetikzlibrary{angles, quotes}
\usepackage{bbm} 
\usepackage{pdfpages}
\usepackage{tabulary}
\usepackage{multirow}
\usepackage{comment}
\usepackage{dsfont}
\usetikzlibrary{calc}
\usepackage{ amssymb }
\usepackage[mathscr]{eucal}
\usepackage{bm}

\renewcommand{\it}{\textit}
\newcommand{\blank}{\:\!}
\usepackage[T1]{fontenc}

\usepackage{multicol}
\usepackage{multirow}
\usepackage{makecell}

\include{notation}
\allowdisplaybreaks 

\begin{document}
\graphicspath{{./Figures/}}
\title{Post-Disaster Communications:\\Enabling Technologies, Architectures,\\and  Open Challenges}
\author{
Maurilio Matracia, \em Student Member, IEEE,
\normalfont Nasir Saeed, \em Senior Member, IEEE,\\
\normalfont Mustafa A. Kishk, \em Member, IEEE, 
\normalfont and Mohamed-Slim Alouini, \em Fellow, IEEE 
\thanks{
Maurilio Matracia and Mohamed-Slim Alouini are with the Computer, Electrical, and Mathematical Sciences and Engineering (CEMSE) Division at King Abdullah University of Science and Technology (KAUST), Thuwal 23955, Makkah, Kingdom of Saudi Arabia (email: {maurilio.matracia; slim.alouini}@kaust.edu.sa);

Nasir Saeed is with the Department of Electrical Engineering, National University of Technology (NUTECH), Islamabad, Pakistan (email: mr.nasir.saeed@ieee.org).

Mustafa Kishk was with KAUST, CEMSE Division during this work. Currently, he is with the Electronic Engineering Department at Maynooth University, Co Kildare, Ireland (email: mustafa.kishk@mu.ie).
}}

\maketitle
\vspace{-1.5cm}

   \begin{abstract}
  The number of disasters has increased over the past decade where these calamities significantly affect the functionality of communication networks. 
      In the context of 6G, airborne and spaceborne networks offer hope in disaster recovery to serve the underserved and to be resilient in calamities.
      Therefore, our paper reviews the state-of-the-art literature on post-disaster wireless communication networks and provides insights for the future establishment of such networks.
      In particular, we first give an overview of the works investigating the general procedures and strategies for facing any large-scale disaster.
      Then, we present technological solutions for post-disaster communications, such as the recovery of the terrestrial infrastructure, installing aerial networks, and using spaceborne networks.
      Afterwards, we shed light on the technological aspects of post-disaster networks, primarily the physical and networking issues.
      We present the literature on channel modeling, coverage and capacity, radio resource management, localization, and energy efficiency in the physical layer part, and discuss the integrated space-air-ground architectures, routing, delay-tolerant/software-defined networks, and edge computing in the networking layer part.
      This paper also includes interesting simulation results which can provide practical guidelines about the deployment of ad hoc network architectures in emergency scenarios. 
      Finally, we present several promising research directions, namely backhauling, cache-enabled and intelligent reflective surface-enabled networks, placement optimization of aerial base stations (ABSs), and the mobility-related aspects that come into play when deploying aerial networks, such as planning their trajectories and the consequent handovers (HOs).
   \end{abstract}


\maketitle
 \vspace{2cm}
\begin{table}[h]  
    \centering
    \caption{Main Nomenclature}
    \begin{tabular}{|l|l||l|l|}
    \hline
\textbf{A2G} & Air-to-ground &
\textbf{ABS} & Aerial base station \\\hline  
\textbf{AI} & Artificial intelligence & 
\textbf{BER} & Bit error rate \\\hline 
\textbf{CSI} & Channel state information &
\textbf{D2D} & Device-to-device \\\hline
\textbf{DTN} & Delay-tolerant network &
\textbf{FR} & First responder   \\\hline
\textbf{FSO} & Free space optics   &
\textbf{HAP} & High-altitude platform \\\hline
\textbf{HO} & Handover &
\textbf{HetNet} & Heterogeneous network \\\hline
\textbf{IoT} & Internet of things &
\textbf{IP} & Internet protocol \\\hline
\textbf{IRS} & Intelligent reflective surface &
\textbf{LAP} & Low-altitude platform   \\\hline
\textbf{LEO} & Low Earth orbit &
\textbf{LiFi} & Light fidelity \\\hline
\textbf{LoRa} & Long range &
\textbf{LoS} & Line-of-sight \\\hline 
\textbf{LTE} & Long term evolution &
\textbf{MANET} & Mobile ad hoc network \\\hline 
\textbf{MDRU} & Movable and deployable resource unit  &
\textbf{MEO} & Medium Earth orbit  \\\hline
\textbf{MIMO} & Multiple-input-multiple-output &
\textbf{NFV} & Network functions virtualization  \\\hline
\textbf{NIB} & Network-in-a-box &
\textbf{NLoS} & Non-LoS \\\hline
\textbf{PDC} & Post-disaster communication &
\textbf{PPDR} & Public protection and disaster relief \\\hline
\textbf{PSO} & Particle swarm optimization &
\textbf{QoS} & Quality of service \\\hline
\textbf{RAN} & Radio access network &
\textbf{RF} & Radio frequency \\\hline
\textbf{S2G} & Space-to-ground &
\textbf{SAGIN} & Space-air-ground integrated network \\\hline
\textbf{SDN} & Software-defined network &
\textbf{SDR} & Software-defined radio \\\hline
\textbf{TBS} & Terrestrial base station &
\textbf{UAV} & Unmanned aerial vehicle \\\hline
\textbf{VANET} & Vehicular ad hoc network &
\textbf{VHetNet} & Vertical HetNet \\\hline
\textbf{WLAN} & Wireless local area network &
\textbf{WSN} & Wireless sensor network 
\\\hline
\end{tabular}
    \label{tab:nomenclature}
\end{table}
\clearpage

\section{Introduction} \label{sec:Intro}
\begin{figure*}
 \centering
    \includegraphics[width=1\textwidth, trim={1.85cm 2.6cm 1.95cm 5.2cm}, clip]{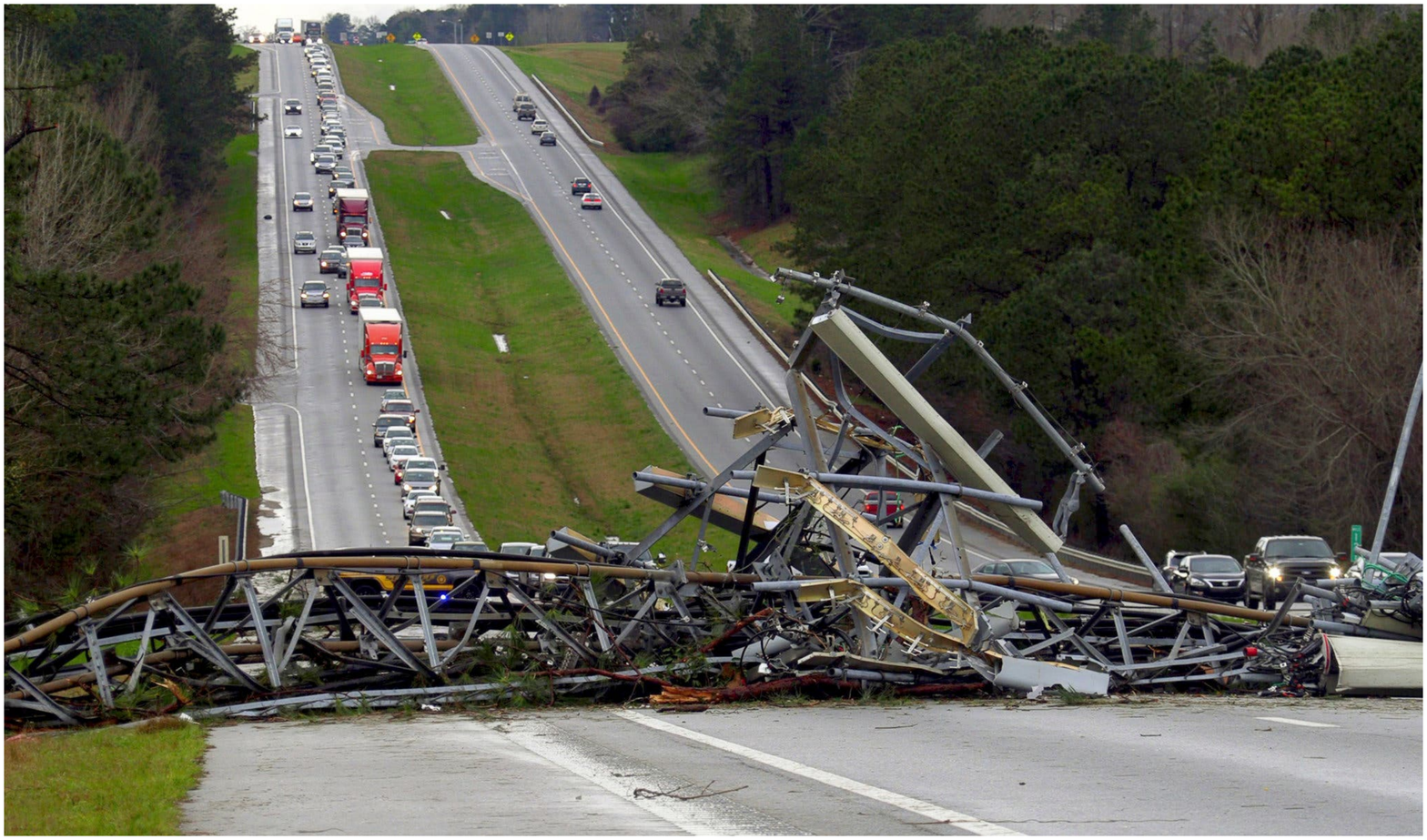}
    \caption{Disaster-struck cell tower~\cite{NYtimes19disasterTower}.}
    \label{fig:disasterTower}
\end{figure*}

While the story of wireless communications tells us the astonishing growth of the achievable data rates over various mobile generations, the increasing research and business interests on the idea of ubiquitous connectivity are much younger. Therefore, wireless communication experts expect that 6G technology will be the first to pay special attention to unconnected and under-connected environments such as low-income, remote, or disaster-struck regions. In this context, many specialized researchers and entrepreneurs are trying to design and implement alternative network architectures and strategies specifically meant to enhance the performances of the current wireless communication systems,  since they are particularly susceptible to calamities (see Fig.~\ref{fig:disasterTower}). \par
In 2021, almost one hundred million people have suffered from natural hazards.
Therefore, several organizations and companies have provided them with tangible support in such circumstances.
For instance, Alphabet's Project Loon, in collaboration with AT$\&$T and T-Mobile, provided connectivity to more than a hundred thousand Puerto Ricans after Hurricane Maria destroyed the local network infrastructure.  
Although the balloons deployed by Alphabet enabled just essential connectivity services, it was an impressive achievement to successfully control their flight from Nevada by using machine learning algorithms~\cite{Loon17PuertoRico}. 
Furthermore, right after the earthquake in Haiti on August 14, 2021, International Telecommunication Union (ITU) and the Emergency Telecommunications Cluster (ETC) collaborated in filling the consequent connectivity gaps experienced by the suffered region.
In particular, they assessed the status of the telecom services via a mapping platform called Disaster Connectivity Map (DCM).
Satellite phones and Broadband Global Area Network (BGAN) terminals were provided by ITU to Haiti~\cite{ITU21haiti}.
Finally, it is worth mentioning Elon Musk's recent efforts in providing a reliable backup \it{Starlink} satellite network to both the Kingdom of Tonga~\cite{reuters22tonga} and Ukraine~\cite{space22musk}, which experienced Internet disruptions due to a tsunami and a conflict with Russia, respectively. \par
Spatial networks (SNs) including non-terrestrial nodes such as ABSs and/or satellites can strongly contribute by providing coverage and capacity at relatively low costs and short deployment times. 
In SNs, the ABSs and satellites operate at very different altitudes; therefore, designing the physical and networking layer is quite challenging. 
For instance, different channel models need to be investigated for characterizing the air-to-ground (A2G), air-to-air, space-to-air, and space-to-ground (S2G) communication. Moreover, energy-related considerations for high-altitude platforms (HAPs) and satellites should be considered since they are mostly solar-powered. \par 
Another crucial aspect to consider is that both power plants and transmission lines are usually susceptible to large scale disasters.
In this context, Ref.~\cite{yuze15autonomous} developed and tested a prototype for an autonomous anti-disaster solar-powered BS connected to the core network via satellite links. \par
Other interesting projects on developing post-disaster communication networks are \it{project Lantern}~\cite{mozilla18lantern}, \it{Baculus}~\cite{mozilla20baculus}, and \it{portable cell initiative}~\cite{mozilla17initiative}.
Project Lantern is a platform based on the long range (LoRa) protocol and aims to support disaster recovery efforts.  It includes a hardware that is able to autonomously provide WiFi, app-based services (essentially meant to help finding supplies), and share useful data to all users through the cloud.
Also, machine learning (ML) is used to make a chatbot assist emergency responders (ERs) by sharing important information.
On the other hand, Baculus creates a WiFi mesh network, where the users are equipped with a so-called divining rod (\it{i.e.}, an antenna that guides them towards the closest WiFi access point). 
The divining rods are continuously updated via satellites about the configuration of the mesh network, in order to help more users to connect.
Alternatively, the portable cell initiative project builds portable and temporary systems acting as cell towers, to be deployed in any unconnected environment. 
In particular, the so-called micro-cells come in a plug-and-play fashion and are characterized by a high level of resilience.
They can interconnect with each other in order to create a mesh network which relies on satellite backhaul in order to provide 2G connectivity, even without new SIM cards. \par
In post-disaster scenarios, connectivity represents a challenging issue and therefore it has gained importance in research over the last few years.
One main reason for this is the difficulty faced when assessing the damages, since a relatively long time is needed to identify the exact zones that lost connectivity due to the calamity.
Moreover, the disaster victims can be trapped in the rubble, making it hard for emergency responders to locate and rescue them.
Several paradigms can be considered for re-establishing connectivity in post-disaster scenarios.
Let us take wildfire detection as an example:
apart from conventional solutions such as satellite imaging and remote camera-based sensing, which are slow and relatively unreliable, the Internet of things (IoT) can be combined with unmanned aerial vehicles (UAVs)~\cite{bushnaq21fire} or even being used in an infrastructure-less manner.
Indeed, IoT-enabled devices can communicate with each other and essentially create a wireless sensor network (WSN) to inform the users about any significant perturbations on their surrounding environment.
To this extent, LoRa technology is often used in order to transfer data over long distances with low power consumption, at the price of a quite limited capacity. 
Moreover, novel drone-assisted mesh network architectures, such as the so-called \it{UbiQNet} ~\cite{ganes21architecture}, are capable of combining image processing and deep learning in order to find the neighbouring first responders (FR) nodes. \par
Some other interesting paradigms that have been explored in the literature in order to support the networking layer are delay-tolerant networks (DTNs) and software-defined networks (SDNs).
The former rely on the \it{Store and Forwarding} routing mechanism for making up to the absence of a direct path between source and destination~\cite{hoque20SDN_DTN}.
SDNs, instead, are meant to decouple forwarding devices' data plane from control plane, in order to ease both the management and the control of the network~\cite{amin18hybrid,lourenco18robust}. 
These two paradigms can be conveniently combined in emergency scenarios. \par
As connectivity is at the heart of humanitarian response in disaster situations, this paper aims to solicit the up-to-date literature on the key aspects of wireless post-disaster communications (PDCs), including both terrestrial and non-terrestrial technologies, as well as the consequent issues involving the physical and the networking layers. 
Furthermore, the proposed work offers relevant simulation results: these can be used by network operators and institutions affected by a calamity in order to estimate the network's performance depending on the considered setup circumstances (essentially, the type and number of available nodes as well as the size of the disaster area).

\subsection{Related Surveys and Reviews}
In order to properly describe the significance of information and communication (ICT) technologies in disaster scenarios, many survey papers have been recently published.\par
Authors in~\cite{nunavath2018role} proposed a systematic review on artificial intelligence (AI) applications for analyzing and processing big data from social media platforms in emergency scenarios. 
Similarly,~\cite{zhao17survey} reviewed social-aware data dissemination approaches and their difference from traditional data dissemination in disaster situations.
In addition, the latest advances regarding the \it{Fog-Assisted Disaster Evacuation Service} (FADE) architecture can be found in Ref.~\cite{Kurniawan21fog}.
\par
The work presented in~\cite{kouyoumdjieva19survey} is an overview of non-image-based techniques for accurately counting people in both indoor and outdoor environments in disaster scenarios. 
The authors of~\cite{wang17hybrid} presented another exciting survey on the role of large-scale 3D networks such as hybrid satellite-aerial-terrestrial networks in emergency scenarios, their architectures, trends, and challenges.
General applications of UAVs in disaster management systems have been extensively reviewed and discussed in~\cite{khan22emerging}. 
A more specific review on UAVs' path planning in smart cities affected by disasters has been recently proposed in~\cite{Qadir21smart}, where several network security aspects are also considered. 
Ref. \cite{recchiuto18assessment}, instead, surveyed UAV-enabled post-disaster assessment, with a special focus on the \it{PRISMA} project, which targets developing and deploying robots and autonomous systems that can monitor and intervene in real-time. \par
A comprehensive review on regulatory and standardization aspects of public safety networks (PSNs) can be found in~\cite{baldini13survey}.
Furthermore, Ref.~\cite{kumbhar2016survey} reviewed emerging paradigms for PSNs,  primarily focusing on how to converge land mobile radio (LMR) and long term evolution (LTE) technologies.
Authors in~\cite{jarwan19lte} also compared LMR and LTE in PSNs, with a special focus on the software environment needed for evaluating the key performance indicators (KPIs).
In a recent survey~\cite{perez20emergency},
the authors discussed future designs for PSNs to manage emergency settings, since in the context of \it{5G and beyond} networks it is required to exploit more advanced technologies such as autonomous decision-making systems and to combine advanced technologies, such as network functions virtualization (NFV) and software-defined networking. \par
Alternatively,~\cite{hwang18multihop} surveyed different multihop ad-hoc network paradigms, including mobile ad hoc networks (MANETs), vehicular ad hoc networks (VANETs), and DTNs, and discussed their importance in the context of disaster response. 
Reviews on routing and rescue entities' mobility models for MANETs in disaster-struck areas can be found in~\cite{Jahir19routing} and~\cite{mahiddin21review}, respectively.
Authors in~\cite{yu18survey} studied disaster recovery solutions from the perspective of users and network solutions such as device-to-device (D2D) and dynamic wireless networks (DWNs), respectively. 
Similarly,~\cite{gomes16survey} presented an overview of the current state of the art of communication systems for mitigating natural disasters, focusing on methods for achieving resilient routing, vulnerability assessment, and reinforcement of existing networks. \par
For non-resilient infrastructures, Ref.~\cite{pozza18network} introduced the network-in-a-box (NIB) concept, which consists in fitting all the required software and hardware modules in portable devices, resulting in a dynamic and versatile architecture which can either self- or inter-operate. 
Authors in~\cite{ali21review} covered the resilience issue as well, showcasing various solutions, including D2D, UAVs, and IoT.
The latter technology is extensively discussed in~\cite{Hasan21search}, where IoT-enabled flood search and rescue (SAR) systems are critically surveyed and a novel IoT-aided integrated flood management framework based on water-ground-air networks is proposed. \par
Authors in~\cite{chiesa21survey} presented a tutorial-like overview of packet-switched networks by focusing on their fast data-plane recovery mechanisms, from traditional layer-2 (\it{i.e.}, data link layer) network technologies to programmable network paradigms and protocols.
Also,~\cite{nazib20routing} presented a state-of-the-art review, categorizing routing protocols for UAV-assisted VANETs based on design and functionality.
In Ref.~\cite{pervez18wireless}, wireless technologies for disaster recovery and healthcare applications were reviewed and compared based on bandwidth, range, and throughput.
{ Contextually, authors in~\cite{carreras22communication} conducted a systematic review on the most recent technologies for emergency communications.
They also introduced a novel architecture, which combines UAV-based wireless mesh networks and near vertical incidence skywave (NVIS) technologies, in order to enable communications between moving agents in harsh environments.}
A summary of the above surveys and reviews is presented in Table \ref{tab:surveys}.

\begin{table}[h]
    \centering
    \caption{Relevant surveys and reviews}
    \begin{tabular}{|c|l|c|}
    \hline
    \textbf{Year} & \textbf{Main focus} & \textbf{Ref.} \\ \hline
    2013 & Regulations and standards &~\cite{baldini13survey} \\ \hline
    \multirow{2}{*}{2016} & Resilience, ad-hoc networks' deployment, routing &~\cite{gomes16survey} \\ \cline{2-3}
    &  LMR-LTE convergence, push-to-talk over LTE  &~\cite{kumbhar2016survey}  \\ \hline
    \multirow{2}{*}{2017} & Hybrid satellite-aerial-terrestrial networks &~\cite{wang17hybrid} \\ \cline{2-3}
    & \makecell[l]{Mobile social networks (MSNs),\\D2D, data dissemination}  &~\cite{zhao17survey} \\ \hline
    \multirow{6}{*}{2018} & Autonomous systems for post-disaster assessment &~\cite{hwang18multihop} \\ \cline{2-3}
    & MANETs, VANETs, DTN &~\cite{hwang18multihop} \\ \cline{2-3}
    & D2D, DWN &~\cite{yu18survey} \\ \cline{2-3}
    & NIB &~\cite{pozza18network} \\ \cline{2-3}
    & AI, big data analysis and processing &~\cite{nunavath2018role} \\ \cline{2-3}
    & \makecell[l]{Wireless technologies for\\disaster recovery and healthcare} &~\cite{pervez18wireless} \\ \hline
    \multirow{3}{*}{2019} & \makecell[l]{Indoor and outdoor non-image-based\\people counting technologies} &~\cite{kouyoumdjieva19survey}\\ \cline{2-3}
    & LMR, LTE &~\cite{jarwan19lte} \\ \cline{2-3}
    & Routing, MANETs &~\cite{Jahir19routing} \\ \hline
    \multirow{2}{*}{2020} & Autonomous decision-making systems, NFV, SDNs &~\cite{perez20emergency} \\ \cline{2-3}
    & UAV-assisted VANETs, routing protocols &~\cite{nazib20routing} \\ \hline
    \multirow{6}{*}{2021} & Wireless technologies for resilient networks &~\cite{ali21review} \\ \cline{2-3}
    & Mobility models, MANETs &~\cite{mahiddin21review} \\ \cline{2-3}
    & Fog computing, disaster evacuation &~\cite{Kurniawan21fog} \\ \cline{2-3}
    & UAV path planning, network security &~\cite{Qadir21smart} \\ \cline{2-3}
    & IoT-enabled flood SAR systems &~\cite{Hasan21search} \\ \cline{2-3}
    & \makecell[l]{Data-plane recovery mechanisms\\for packet-switched networks} &~\cite{chiesa21survey} \\ \hline
    \multirow{3}{*}{2022} & UAVs for disaster management & \cite{khan22emerging} \\\cline{2-3}
    & Technologies for emergency communications & \cite{carreras22communication} \\\cline{2-3}
    & \makecell[l]{Paradigms, physical and networking layers, challenges} & \makecell[c]{This\\paper} \\ \hline
    \end{tabular}
    \label{tab:surveys}
\end{table}

\subsection{Contributions}
Based on the relevant surveys discussed above, and to the best of the authors' knowledge, there is no single survey that distinctly focuses on the updated communication technologies and issues in post-disaster situations. 
Since communication is a prominent part of establishing post-disaster networks, it is pertinent to analyze the connection between disaster situations and communication technologies from various aspects, including technical issues, applications, and challenges. Therefore, this paper offers an overview of the main features of PDCs, with particular attention to ad-hoc aerial networks. We can thus summarize our contributions as follows: 
\begin{enumerate}
    \item We provide an up-to-date review of PDCs to project the reader in the perspective of the 5G and beyond generation of wireless communications. 
    In particular, we extensively discuss the most relevant research efforts done over the last decade to design powerful wireless technologies for PDCs and try to solve recurrent physical and networking layer problems;
    \item We present our stochastic-geometry-based simulation results for two realistic post-disaster network setups, and consequently discuss how to achieve efficient network planning in such scenarios;
    \item Given the numerous aspects deserving further researchers' attention, we discuss inherent challenges and present what we believe are the most promising directions for future research in the area of post-disaster wireless communications. 
    In particular, our interest is focused on: modulation and coding, backhauling, placement, trajectory and scheduling of movable nodes, HO management, and content caching.
\end{enumerate}

\subsection{Organization}

\begin{figure*}
 \centering
    \includegraphics[width=1\textwidth, trim={9.6cm 0.5cm 0.7cm 0.5cm}, clip]{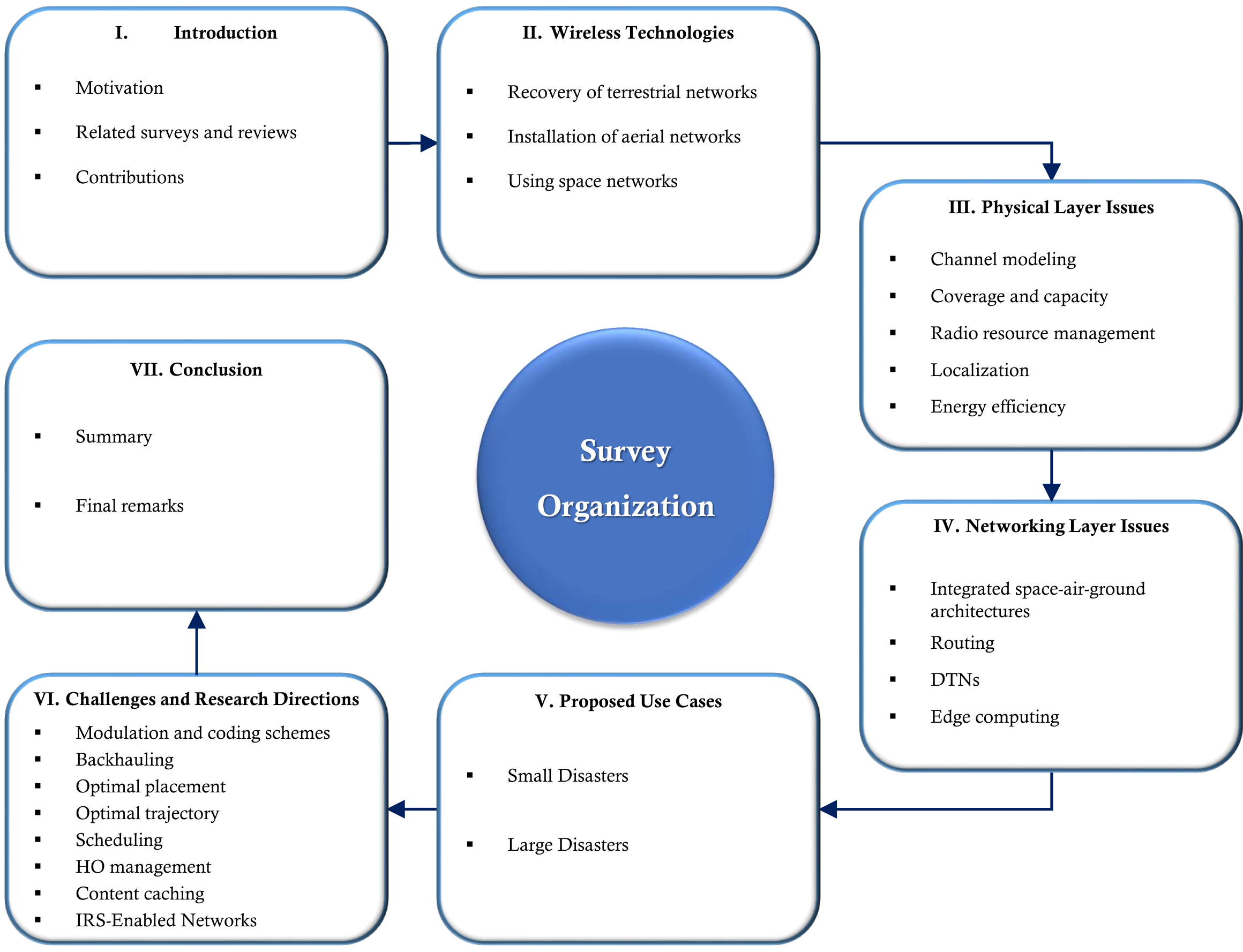}
    \caption{{Structure of this paper.}}
    \label{fig:diagram}
\end{figure*}

The rest of the paper is organized as follows. 
Sec.~\ref{sec:wirelessTechs} discusses the most relevant wireless technologies for disaster management and recovery.
In Sec.~\ref{sec:physicalLayer}, we review the main physical layer issues, with a particular focus on the most common channel models as well as the techniques for improving energy efficiency in emergency scenarios.
Nonetheless, other literature works referring to inherent aspects such as coverage, capacity, radio resource management, and localization are extensively discussed.
In addition, the same section offers two realistic use cases about wireless coverage in disaster environments and exploit a stochastic geometry approach for our numerical simulations.
On the other hand, Sec.~\ref{sec:networkingLayer} provides an overview of some recent works  on the networking layer aspects of integrated space-air-ground architectures, routing, delay-tolerant networking, and software-defined networking.
Finally, Sec.~\ref{sec:Challenges} shows the main future research challenges, followed by our conclusions in Sec.~\ref{sec:conclusion}.\par
The structure of this paper is schematically displayed in Fig.~\ref{fig:diagram}.

\section{Wireless Technologies} \label{sec:wirelessTechs}
\begin{figure*}
 \centering
    \includegraphics[width=1\textwidth, trim={0.02cm 0cm 0cm 0cm}, clip]{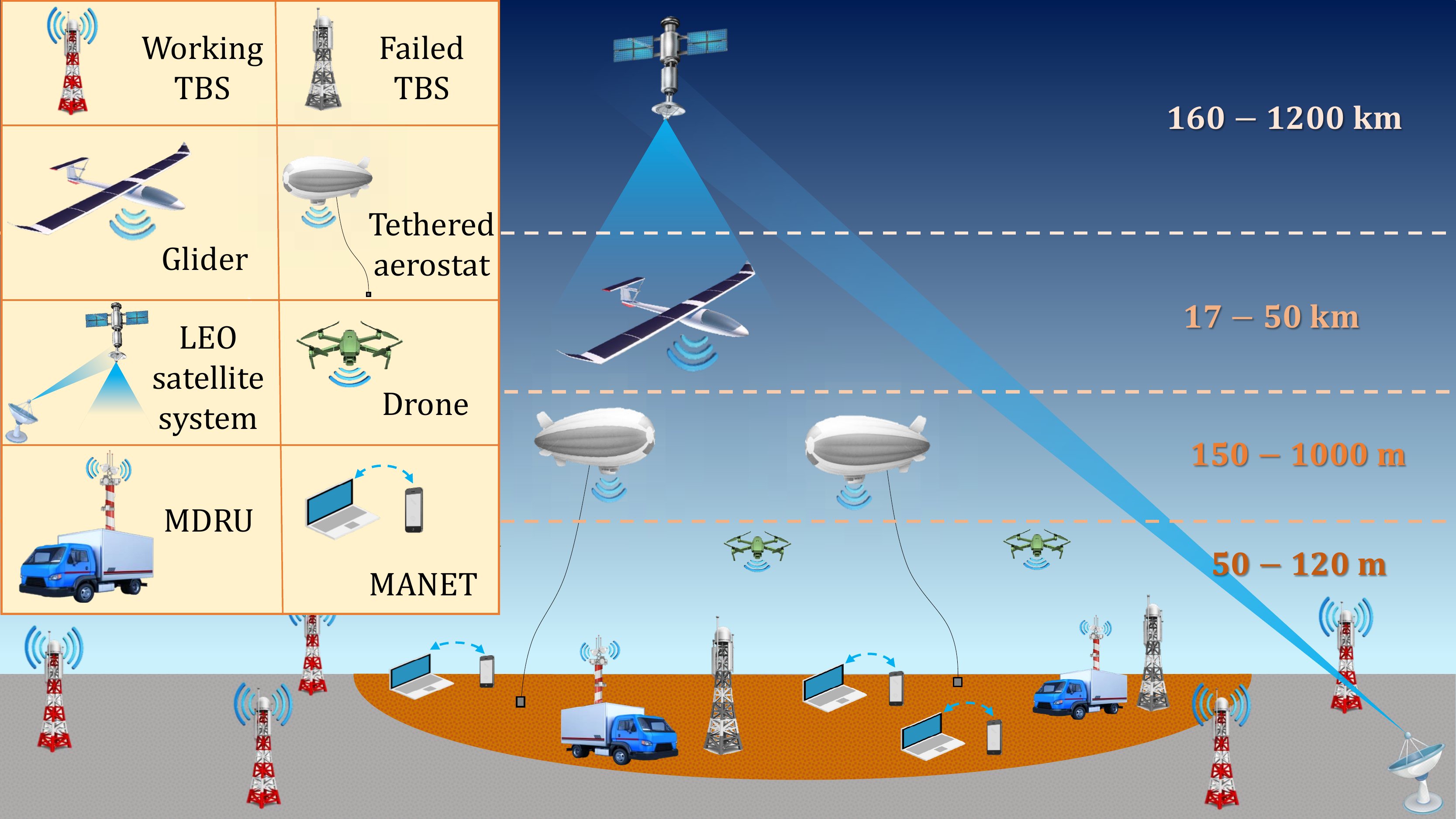}
    \caption{Integrated communication systems including the main ad hoc network paradigms, namely VANETs, MANETs, LAPs (\it{e.g.}, drones, tethered aerostats), HAPs (\it{e.g.}, gliders), and low Earth orbit (LEO) satellite systems. The area in orange represents the disaster-struck zone.}
    \label{fig:arch}
\end{figure*}

\begin{figure}
 \centering
    \includegraphics[width=0.6\columnwidth, trim={8cm 0.4cm 7.2cm 0.5cm}, clip]{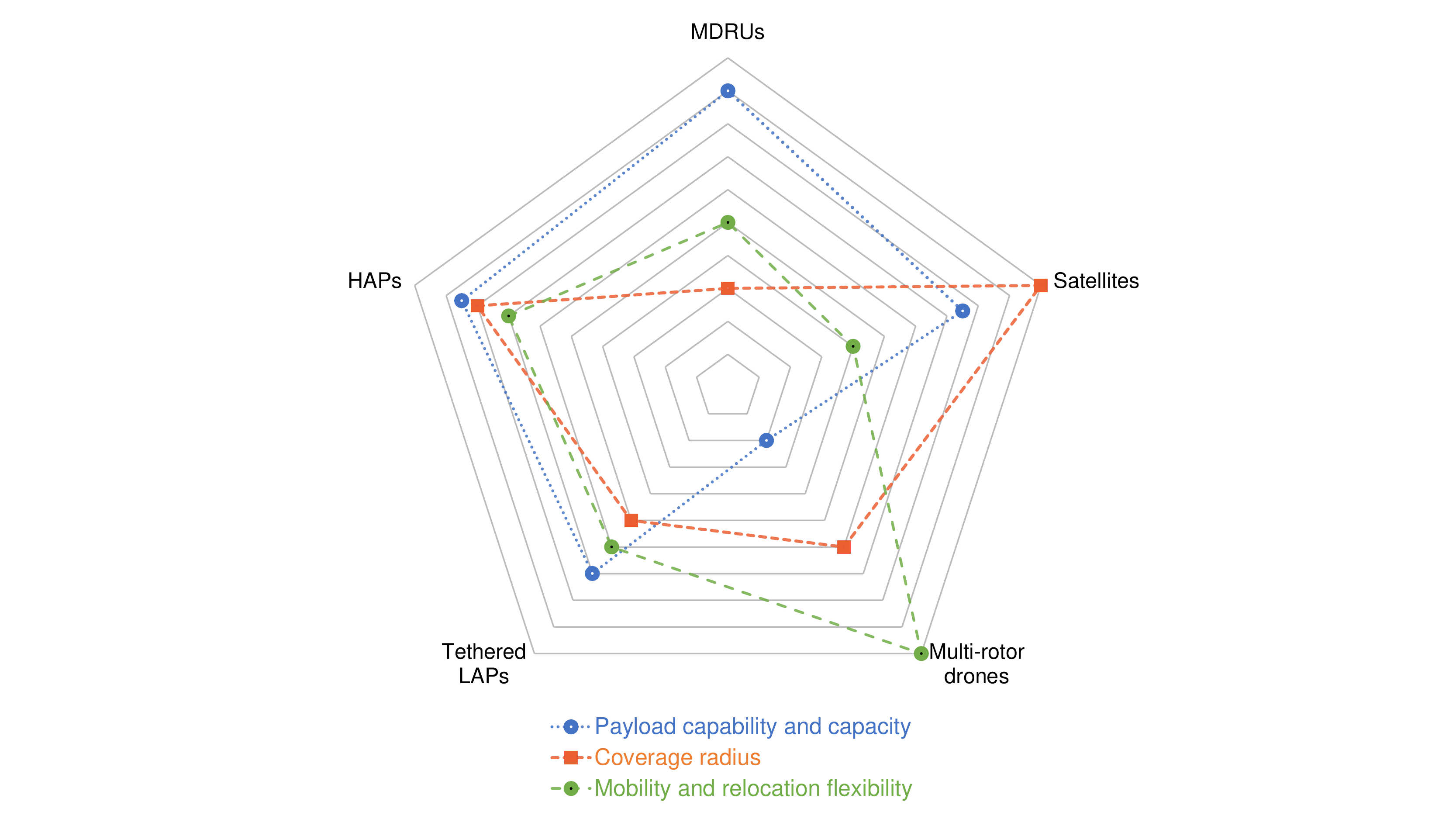}
    \caption{ Qualitative comparison between various types of platforms.}
    \label{fig:radar_platforms}
\end{figure}

In post-disaster situations, a timely counteraction is generally required.
Therefore, many wireless technologies have been specifically designed or even re-adapted for serving victims and FRs.
In this section, we discuss literature works that recently proposed solutions for improving PDCs, and categorize them depending on the network architecture (namely terrestrial, aerial, or space enabled networks).
A schematic view of a comprehensive network architecture for PDCs is illustrated in Fig.~\ref{fig:arch}, while Fig~\ref{fig:radar_platforms} compares various cellular paradigms.

\subsection{Recovery of Terrestrial Networks}
One main limitation of current network infrastructures is the lack of sufficient redundancy.
Moreover, the most recent occurrences of calamities have often demonstrated that wireless networks are very susceptible. 
Whenever the terrestrial infrastructure gets damaged, it is preferable to consider repairing it before resorting to other ad-hoc solutions such as ABSs.
{However, it is evident that several challenges may arise:
tight time constraints, scarce availability of specialized installers and respective equipment, and harsh environmental conditions are just few examples of the difficulties to be potentially faced.} \par
In this context,  the authors of~\cite{deepak19robust} have analyzed the recovery phase of a communication network, evaluating the advantages of D2D and cellular communication systems operating in underlay mode.
Furthermore, Ref.~\cite{deepak2019overview} has investigated the three possible disaster-struck network conditions, namely congested, partly functional, or fully isolated, including important considerations on spectrum allocation. Contextually, authors in~\cite{udreakh18RAPID} have promoted \it{RAPID TIMEER}, a system for recording and reporting (via texts, images, and voice) that works independently from both power and telecommunication infrastructures. \par
The authors of~\cite{narang17cyber} propose the use of cyber-physical systems, especially in public buses and drones, to develop a mobile edge infrastructure, where the buses host as BSs, computation units, and power resources, and are can thus support drones in covering hard-to-reach areas. \par
{Park \it{et al}.~\cite{park21demand} pushed the NIB concept forward by proposing their vision in the context of 6G, and the consequent challenges to overcome.
The authors also proposed an interesting case study where NIB's flexibility is exploited to maximize spectral efficiency in maritime communications.} 
Finally, Sakano \it{et al}.~\cite{sakano16bringing} suggested deploying so-called movable and deployable resource units (MDRUs)\footnote{$\,$Although these units are typically terrestrial base stations, in this paper the acronym `TBS' exclusively refers to any node of the fixed terrestrial infrastructure, such as a cell tower.}, which are essentially vehicle-mounted base stations, as a comprehensive solution for satisfying the needs of users in disaster recovery situations.
Some of the main advantages of this solution are indeed represented by its agility, prompt installation, and carrier-free usability.

\subsection{Installation of Aerial Networks}
Since the recovery of terrestrial networks in disaster-affected regions is usually a long process, many works focused on quicker solutions such as using aerial networks, with airborne platforms (drones, balloons, gliders, airships, \it{etc.}) working as flying BSs. Depending on the height and resources of the aerial BS, we can broadly categorize them as low-altitude platforms (LAPs) or HAPs. \par
{ The main challenges associated to aerial networks in disaster scenarios are mostly related to their deployment and operation.
Indeed, advanced technologies are mostly required for: timely moving and deploying the aerial nodes, controlling their trajectory (especially in case of swarms of untethered drones, as will be further discussed in Sec.~\ref{subsec:traj}), ensuring sufficient flight endurance (which relates to the problem of energy efficiency discussed in Sec.~\ref{subsec:en_eff}).
Depending on the altitude of the aerial node, different types of aerial networks can be used for various disaster management applications and use cases~\cite[Chapter 9]{bushnaq22unmanned}.
}

\subsubsection{LAP-Based Solution}
Due to their fast implementation, a large number of works discussed LAPs usage in PDCs.
For example, Ref.~\cite{Erdelj16_disasterManagement} reviewed disaster management applications and challenges using UAV networks, especially when combined with WSNs and cellular networks.
Similarly, authors in~\cite{Naqvi18key} extensively discussed UAV-aided disaster-resilient networks from a 5G perspective, including fruitful considerations on the simultaneous occurrence of UAV and D2D communications and the power control strategies. 
Furthermore,~\cite{Matracia21disaster} recently presented the main topological elements to take into account when deploying ABSs, where various types of UAVs are compared for typical post-disaster network setups.
Authors in~\cite{mase15message}, instead, introduced a distributed and scalable message-based system relying on electric vehicles (EVs) and UAVs that allows to connect shelters by properly partitioning the suffered region (assuming it is sufficiently large).
Finally, the main targets of UAV-aided networks, namely ubiquitous coverage, relaying, and information dissemination, were extensively discussed in~\cite{Zeng16unmanned} together with some design considerations and performance-enhancing techniques. \par
A recent trend shows a considerable interest in using optimization tools for aerial networks-assisted post-disaster communications.
For example, authors in~\cite{li18UAV} used particle swarm optimization (PSO) to adjust UAVs' antenna altitude and beam angle for post-disaster networks. 
The developed framework allows to effectively optimize coverage under the transmit power constraint.
Similarly, authors in~\cite{do-duy21joint} proposed another optimization-based work for UAV-aided disaster communications (\it{i.e.}, a macro BS supported by UAVs serving hard-to-reach clusters of users), and validated their framework for ensuring high energy efficiency by jointly optimizing UAVs' deployment and resource allocation. 
Then, Ref.~\cite{Zhao19} proposed a unified framework that considers UAVs' trajectory, scheduling, and transceiver design optimization in case of emergency. 
Similarly, authors in~\cite{Merwaday16} used a genetic algorithm (GA) to evaluate improvements in terms of throughput when the flying BSs are optimally placed; in particular, the study showed that the performance of the network can generally be improved by increasing the number of ABSs and decreasing their altitude.
Also, Ref.~\cite{zobel20topology} recently presented a novel approach to rapidly detect users' clusters variations in a post-disaster situation, where data ferry UAVs' path planning is contextually optimized to connect the highest number of nodes in a reasonable amount of time.
Furthermore, authors in~\cite{xu20time} proposed an integrated aerial-ground network for swift communication recovery to maximize the time-weighted coverage (\it{i.e.}, the integration over time of the coverage area weighted by a time-dependent function) for a given deployment strategy. \par
The combined use of ABSs and D2D technologies can extend the coverage for PDCs. 
For instance, authors in~\cite{miao20joint} proposed a linear programming algorithm for obtaining a suboptimal solution for the problem of maximum rate and coverage of UAV-enabled networks with underlying D2D communications.
However, the latter work neglects the effects of both small-scale fading and non-line-of-sight (NLoS) transmission.
In the same context,~\cite{liu19transceiver} introduced two optimal transceiver designs schemes and a shortest-path-routing algorithm to construct efficient multihop D2D links in a post-disaster situation. \par
Besides the aforementioned physical layer issues, various works tackled the network layer issues in PDCs.
For example, ~\cite{grover17operated} discussed the layout of the aerial emergency ad-hoc network (EANET) for both ad-hoc on demand distance vector (AODV) and zone routing protocol (ZRP), eventually promoting the latter since it generally leads to a higher packet delivery fraction (\it{i.e.}, the ratio between the number of packets generated at the source and the number of packets received at the destination). 
Furthermore, the study in~\cite{solpico19UAV_DTN}  suggested a UAV-DTN based on a decentralized near-cloud infrastructure with LoRa technology to provide low-power transmission over long distances. The authors investigated the performances of the network (\it{i.e.}, long-range detection/messaging and detection under rubble) in both inside and outside environments, in order to comprehensively test the proposed technology.
Another work~\cite{mori15construction} recently introduced a UAV-assisted DTN that takes into account the importance of UAVs' altitude, since it impacts the required number of nodes, delay time, and delivery ratio. 
This work is particularly interesting because it takes into account also the type of environment (urban, suburban, or rural).

\subsubsection{HAP-Based solution}
HAPs can be either helikites, airships, gliders, or balloons operating at higher altitudes than LAPs, and their  characteristics are intermediate between drones and satellites.
Their main advantage is probably the much longer endurance, which implies they could even support PDCs without being deployed after the occurrence of a calamity: HAPs can indeed be aloft for several months, and provide coverage and capacity even in ordinary conditions of the network, meaning that they are not necessarily an ad hoc solution.
On the other hand, the time they require for proper deployment and relocation (they generally lack a propulsion system) might be excessive for most of the emergency situations.\par
Authors in~\cite{gomez13realistic} considered helikites to deploy 4G-LTE remote radio head (RRH) and provide high capacity Internet services in case of emergency.
Alternatively, other works use balloons in disaster-struck areas (recall Google Loon in Puerto Rico) to quickly recover the networks.
For instance, the work presented in~\cite{Surampudi2018a} proposed a novel network of balloons equipped with light fidelity (LiFi) transceivers, with a particular focus on the physical design of the platforms. Moreover, the authors of~\cite{arimura14micro} propose an interesting wireless balloon monitoring system, where a high-resolution omnidirectional camera and wireless LAN technologies are exploited to make bird's eye views available to ground relay stations. On the other hand, Ref.~\cite{nakajima15relay} introduces a satellite-aided balloon-based wireless relay system which can be deployed in less than four hours.\par
 Besides floating balloons, there is a recent trend for using tethered balloons to enable connectivity in disaster situations. Authors in~\cite{alsamhi2018} presented a comprehensive overview of the characteristics of tethered balloons.
 More recently, Ref.~\cite{rengaraju21balloon} compared various technologies and promoted the use of WiFi balloon to access social networks in the occurrence of a calamity.
 \par
Finally, the work presented in~\cite{gharbi19crises} tackled important challenges such as routing and resource allocation for heterogeneous vertical networks that include not only HAPs, but also satellites and cell towers.
A summary of the papers mentioned in this section can be found in Table \ref{tab:wirelessTech}.
    
    \begin{table*}[]
    \centering
    \caption{Summary of the relevant papers on wireless technologies}
    \begin{tabular}{|c|c|l|c|}
    \hline
    \textbf{Topic} & \textbf{Year} & \textbf{Main focus} & \textbf{Ref.} \\ \hline
    \multirow{5}{*}{Recovery of terrestrial networks} & 2016 & MDRUs &~\cite{sakano16bringing}  \\ \cline{2-4}
    & 2017 & Mobile edge infrastructure, cyber-physical systems &~\cite{narang17cyber} \\ \cline{2-4}
    & 2018 & Disaster assessment, data inventory, field situation recording/reporting &~\cite{udreakh18RAPID} \\ \cline{2-4}
    & \multirow{2}{*}{2019} &  D2D and cellular systems &~\cite{deepak19robust}  \\ \cline{3-4}
    & & 4G-LTE, D2D, UAVs, MANETs, IoT &~\cite{deepak2019overview} \\\cline{2-4}
    & 2021 & 6G-NIB, maritime communications &~\cite{park21demand}\\ \hline
    \multirow{23}{*}{Installation of aerial networks} & 2013 & Helikites, 4G-LTE RRH &~\cite{gomez13realistic} \\ \cline{2-4}
    & 2014 & Balloons, monitoring &~\cite{arimura14micro} \\ \cline{2-4}
    & \multirow{3}{*}{2015} & EVs, UAVs, area partitioning &~\cite{mase15message} \\ \cline{3-4}
    & & DTN, topology, altitude &~\cite{mori15construction}  \\ \cline{3-4}
    & & Balloons, satellites, relaying, deployment time &~\cite{nakajima15relay} \\ \cline{2-4}
    & \multirow{3}{*}{2016} & Coverage, relaying, information dissemination &~\cite{Zeng16unmanned} \\ \cline{3-4}
    & & UAVs, WSNs &~\cite{Erdelj16_disasterManagement}  \\ \cline{3-4}
    & & Optimal placement, throughput &~\cite{Merwaday16}  \\ \cline{2-4}
    & 2017 & EANET, AODV, ZRP &~\cite{grover17operated} \\ \cline{2-4}
    & \multirow{4}{*}{2018} & Balloons, LiFi &~\cite{Surampudi2018a} \\ \cline{3-4}
    & & Tethered balloons & ~\cite{alsamhi2018} \\ \cline{3-4}
    & & UAVs, resilience, D2D, power control & ~\cite{Naqvi18key} \\ \cline{3-4}
    & & Implementation of PSO algorithm & ~\cite{li18UAV} \\ \cline{2-4}
    & \multirow{4}{*}{2019} & Trajectory, scheduling, transceiver design &~\cite{Zhao19} \\ \cline{3-4}
    & & Transceiver design, routing, D2D &~\cite{liu19transceiver} \\ \cline{3-4}
    & & Heterogeneous networks (HetNets), routing, resource allocation &~\cite{gharbi19crises} \\ \cline{3-4}
    & & DTN, LoRa &~\cite{solpico19UAV_DTN} \\ \cline{2-4}
    & \multirow{3}{*}{2020} & Cluster localization, UAV path planning &~\cite{zobel20topology} \\ \cline{3-4}
    & & Swift communication recovery, time-weighted coverage &~\cite{xu20time} \\ \cline{3-4}
    & & Rate, coverage, D2D &~\cite{miao20joint}  \\ \cline{2-4}
    & \multirow{3}{*}{2021} & WiFi balloons, social network &~\cite{rengaraju21balloon} \\ \cline{3-4}
    & & Deployment, resource allocation &~\cite{do-duy21joint} \\ \cline{3-4}
    & & Topological aspects, capacity, types of fleets &~\cite{Matracia21disaster}  \\ \hline
    \multirow{7}{*}{Using space networks} & \multirow{3}{*}{2013} & LTE-satellite and radio interface &~\cite{syiang13chinese} \\ \cline{3-4}
    & & Satellite-gateway links and WINDS &~\cite{takahashi13WINDS} \\ \cline{3-4}
    & & SDR-VSAT systems and coverage &~\cite{suematsu13VSAT} \\ \cline{2-4}
    & 2015 & Satellite-LTE integration for disaster recovery &~\cite{casoni2015integration} \\\cline{2-4}
    & \multirow{2}{*}{2019} & IoT and big data analytics &~\cite{lwin19geospatial} \\ \cline{3-4}
    & & Weather prediction and evacuation warning systems &~\cite{salvi19weather}  \\ \cline{2-4}
    & 2021 & Experimental field trials on novel satellite-based architectures &~\cite{volk21satellite} \\\hline
    \end{tabular}
    \label{tab:wirelessTech}
\end{table*}
    
\subsection{Using Space Networks}
Last but not least, emerging small satellite networks can play a major role in providing connectivity to the disaster struck regions~\cite{Nasir2021}. \par
{ Although the main use of satellites in post-disaster scenarios consists in providing backhaul to the aerial nodes (especially HAPs, as we assumed in Sec,~\ref{subsubsec:large_disast}), one interesting research direction is to design low-latency-LEO satellites that are able to provide also access functionalities to the users, similarly to ABSs.
However, power consumption is still one of the many major issues~\cite[Sec. IX-D]{kodheli21satellite}.}
Nonetheless, several works have considered the use of satellites in disaster situations over the last decade~\cite{volk21satellite,casoni2015integration,syiang13chinese,takahashi13WINDS,lwin19geospatial,salvi19weather,dai20constellation,suematsu13VSAT,shin17mixed}.\par
{
For example, Ref.~\cite{volk21satellite} presented three novel architectures, supported by experimental measurements, based on satellite communication (SATCOM) for public protection and disaster relief (PPDR).
The proposed schemes (namely, SATCOM-enabled network with two independent evolved packet cores, network with synchronized evolved packet cores and multi-user-equipment router, and hybrid SATCOM-enabled network with the multi-user-equipment router) have different pros and cons, all aiming to minimize the network deployment time while maintaining the network secure and versatile.
In the same context, a combination of Ka-band medium Earth orbit (MEO) satellites (for backhaul) and LTE (for access) was proposed in ~\cite{casoni2015integration}.
In particular, their work suggested infrastructure-less and infrastructure-based topologies relying on ad hoc mobile emergency operations control centers (MEOCs) to provide connectivity to FRs.} \par
Authors in~\cite{syiang13chinese} introduced the Chinese initiative called LTE-satellite and comprehensively discussed its radio interface technology.
Then, in~\cite{takahashi13WINDS}, the authors presented an experimental work based on \it{Wideband Internetworking engineering test and Demonstration Satellite} (WINDS), showing several examples of satellite-earth station links. Due to the importance of geospatial data for IoT-aided disaster management, Ref.~\cite{lwin19geospatial} discussed the so-called \it{City Geospatial Dashboard} which can be utilized for collecting, sharing, and visualizing geospatial big data. \par
Other interesting works such as~\cite{salvi19weather} focused on the use of satellites in the context of natural disasters, specifically for weather prediction and warnings for easier evacuation.
Finally,~\cite{suematsu13VSAT} discussed multi-mode software-defined radio (SDR) for very small aperture terminal (VSAT) systems as a possible solution to provide cellular coverage in case of large scale disasters, such as the great East Japan earthquake occured in early 2011.

\subsection{Important Remarks}
{ Assuming a moderate severity of the calamity, increased network redundancy (whenever economically feasible) may be the best solution to prevent outages in post-disaster scenarios. 
This is because if the infrastructure gets damaged it may be difficult to repair or support it in a timely manner. 
Alternatively, there are solutions such as airborne and spaceborne networks; despite they still present several techno-economic limitations, their deployment has been quite effective in several cases.
Other paradigms such as NIB and MDRU also have the potential to overcome this issue, at least in case of small disasters.\par
Nonetheless, it is evident that disasters are often unpredictable and sometimes governments and network operators are not prepared to face them.
Therefore, important efforts are required to improve at least the general emergency plans and ensure a decent level of network resilience: while this might not lead to the optimal strategies, it will ensure timely counteractions to any calamities since all wireless technologies can be helpful in challenging situations. 
}

\section{Physical Layer Issues} \label{sec:physicalLayer}
In this section, we survey the main topics related to the physical layer of a post-disaster network.
We will mostly focus on the differences between the channel models used for terrestrial, aerial, and space communications.
However, to the best of our knowledge literature works considering degradation of the environment, for example due to the smoke generated by a wildfire or the debris brought by a tornado, are still missing. \par
Another important issue that will be covered in this section is related to coverage and capacity in PDCs. 
These two performance metrics gain great importance in critical situations, since what often saves trapped victims' lives is being able to access emergency information as well as to share their location to the outer rescuers, which explains why also the problem of localization is included hereby.

\subsection{Channel Modeling} \label{subsec:channelMod}
{ Post-disaster environments can be particularly harsh because of the high degree of scattering due to the presence of obstacles such as rubble, fallen trees, ash, \it{etc}.}
Therefore, channel modeling is one of the major aspects of post-disaster communication networks, especially when considering vertical heterogeneous networks (VHetNets) since the complexity of the network topology is considerably high.
Due to this, the general fading distributions (\it{e.g.}, Rayleigh, Rician, and Nakagami-\it{m}) are not always applicable in post-disaster scenarios~\cite{yao21resource}.
In this subsection, we survey the literature about PDCs' channel modeling for non-terrestrial networks. 
A summary of the reviewed papers can be found in Table~\ref{tab:channel}.

\subsubsection{LAPs} \label{subsubsec:chann_LAPs}
In~\cite{al2014optimal} we can find an accurate mathematical model allowing us to estimate the ABS' altitude that maximizes the coverage area and a closed-form expression for computing the line-of-sight (LoS) probability. 
Then, the authors of~\cite{AlHourani14b} provided a statistical propagation model for predicting the A2G path loss between terrestrial and aerial nodes, given the urbanization level of the environment and the ABS' elevation angle.
In the latter paper, reflections due to objects and trees where neglected for simplicity, while the one due to buildings was modeled under the assumption that their surface was made of concrete, which has considerable dielectric parameters leading to strong reflection phenomena;
also, the authors assumed knife-edges in order to evaluate scattering in a deterministic manner, although this implies approximated results. \par 
Furthermore, it should be noted that for urban environments a large percentage of the victims is usually trapped inside buildings with multiple floors, which requires to use a tridimensional model.  
Therefore, Ref.~\cite{ranjan18pathloss} compared the performances of various path loss models as well as the impact of indoor and outdoor environments on both uplink and downlink, which in general are not symmetrical.
To model LoS propagation channels usually the Winner II and the free-space pathloss models are used, whereas for the case NLoS propagation the majority of the works relies on Winner II and two-ray models.
Hence, the authors of~\cite{ranjan18pathloss} investigated the Winner II pathloss model proposed in~\cite{meinila2009winner} with an extra blockage component (which refers to the indoor part of the path), and found it to be the most accurate for urban environments. \par
Finally, a recent work~\cite{yao21resource} presented a novel framework to characterize the composite fading channel and optimize both capacity and energy efficiency.
In particular, the authors used the Fisher-Snedecor $\mathcal{F}$ distribution to characterize the link between UAVs equipped with intelligent reflective surfaces (IRSs) and trapped users, and proved the effectiveness of their resource allocation scheme by means of selected simulation results.

\subsubsection{HAPs} 
To the best of our knowledge, there are no works exclusively focusing on HAPs' channel modeling for PDCs.
However, several contributions can be extracted from relevant references with a general validity, as discussed in what follows.\par
Ref.~\cite{iskandart17LTE} assumed a Rician fading channel with K factor for analyzing capacity in HAP networks.
The choice of using the Rician fading channel model can be justified by noting that HAPs have a wide elevation angle, which allows them to be almost always in LoS conditions with the typical user within the same cell.
Authors in~\cite{yang17statistical} statistically modeled the HAP dual circularly polarized 2$\times$2 multiple-input-multiple-output (MIMO) propagation channel and applied the ray tracing approach to the digital relief model to solve the problem of lacing measured data.
Furthermore, a theoretical 3D wideband model was introduced in~\cite{lian17wideband} for HAP-MIMO channel. 
Note that the tridimensional model was needed because of the considerable altitude of the flying base station, which was assumed to be equipped with multiple transmit and receive antennas aligned in different planes.
In the latter work, the Chapman-Kolmogorov equations were applied in order to derive the survival probabilities of scatterers.
motivated by the absence of experimental results in the literature, authors in~\cite{zhao20Kaband} presented a novel statistical channel model of long-distance \it{Ka-band} signal transmission via HAP and verified it via numerical simulations.
Finally, authors in~\cite{safi20analytical} derived tractable closed-form statistical channel models for ground-to-HAP free-space optics (FSO) links which also take into account the effects of atmospheric turbulence and other relevant aspects of FSO communications.

\subsubsection{Satellites} 
For space communications, it is still needed to model the S2G channel because of the current lack of achievable standards.
Indeed, even if some standards were established by the Consultative Committee for Space Data Systems (CCSDS), they are often obstructed by technological limitations~\cite{saeed20cubesat}.
The communication link could leverage either laser, radio frequency (RF), or visible light communication (VLC).
The critical aspect to take into account is the energy consumption for each technique, since satellites essentially rely on solar energy. \par
The signal envelope is subjected to three main sources of variation, namely multipath fading, LoS shadowing, and multiplicative shadow fading.
The first one, which is usually modeled by Rayleigh or Rice distributions, is generated by the combination of all the scattered NLoS components along with a possible LOS ray, leading to rapid small-scale fluctuations. 
On the other hand, LoS shadowing arises from a LoS obstruction due to objects (\it{e.g.}, trees and buildings), which implies slower fluctuations on a larger scale.
Finally, the multiplicative shadow fading phenomenon  is responsible for random variations in the power of LoS multipath components. 
Most of the famous models consider land-mobile-satellite communication systems, and can be categorized as static (\it{e.g.}, the ones developed by Loo~\cite{loo85statistical}, Corazza-Vatalaro~\cite{corazza94statistical}, Hwang~\cite{hwang97channel}, Patzold~\cite{patzold98study}, Kourogiorgas~\cite{kourogiorgas14modeling}, Abdi~\cite{abdi03new}, and Saunders~\cite{saunders96physical}) or dynamic (\it{e.g.}, the ones developed by Fontan~\cite{fontan01statistical}, Scalise~\cite{scalise06accurate}, Nikolaidis~\cite{nikolaidis16dual}, and Lopez-Salamanca~\cite{lopez19finite}). 
\par
Authors in~\cite{aparna17parametric} considered a dual-polarized MIMO channel, focusing on the models referred to as \it{Quasi Deterministic Radio channel Generator} (QuaDRiGa) and \it{Loo} in order to capture the ionospheric, tropospheric, and fading effects on a land mobile satellite (LMS) system.
In the same context, authors in~\cite{wang17algorithm} introduced an algorithm for modeling dual-polarized MIMO channel while taking into account LoS shadowing, multipath effect, elevation angle, and other relevant channel factors.
Ref.~\cite{alSaegh14analysis} introduced a reliable channel model for taking into account dynamic cloudy weather impairments, which impact on the Rician factor and signal propagation.
On the other side, a simplified fading channel model for describing received signals, multipath fading, and shadowing effect has been analyzed by the authors of Ref.~\cite{liu16analysis}.
In~\cite{zou19dynamic}, instead, the so-called channel reservation strategy is promoted as a solution for improving access and HO performances.
Finally, authors in~\cite{bai20novel} presented an original channel model for 5G and beyond which, based on atmospheric data, allows to predict channel attenuation at any time.

\begin{table}[]
    \centering
    \caption{Summary of the relevant papers on channel modeling}
  \begin{tabular}{|c|c|l|c|}
    \hline
    \textbf{Topic} & \textbf{Year} & \textbf{Main focus} & \textbf{Ref.}  \\ \hline
    \multirow{5}{*}{LAPs} & \multirow{2}{*}{2014} & Altitude, LoS probability &~\cite{al2014optimal} \\ \cline{3-4}
    & & A2G path loss &~\cite{AlHourani14b} \\ \cline{2-4}
    & \multirow{2}{*}{2018} & Resilience &~\cite{Naqvi18key} \\ \cline{3-4}
    & & Path loss &~\cite{ranjan18pathloss} \\ \cline{2-4}
    & 2021 & Power allocation &~\cite{yao21resource} \\ \hline
    \multirow{5}{*}{HAPs} & \multirow{3}{*}{2017} & Rician fading channel, capacity &~\cite{iskandart17LTE} \\ \cline{3-4}
    & & MIMO, ray tracing &~\cite{yang17statistical} \\ \cline{3-4}
    & & MIMO, Chapman-Kolmogorov equations &~\cite{lian17wideband} \\ \cline{2-4}
    & \multirow{2}{*}{2020} & FSO &~\cite{safi20analytical} \\ \cline{3-4}
    & & Ka-band transmission &~\cite{zhao20Kaband} \\ \hline
    \multirow{5}{*}{Satellites} & 2014 & Rician factor, signal propagation &~\cite{alSaegh14analysis} \\ \cline{2-4}
    & 2016 & Fading channel &~\cite{liu16analysis} \\ \cline{2-4}
    & 2017 & Dual-polarized MIMO, LMS & \makecell{\cite{aparna17parametric},\\\cite{wang17algorithm}} \\ \cline{2-4}
    & 2019 & Channel reservation strategy, HO &~\cite{zou19dynamic} \\ \cline{2-4}
    & 2020 & Channel attenuation &~\cite{bai20novel} \\ \hline
\end{tabular}
\label{tab:channel}
\end{table}

\subsection{Coverage and Capacity}
This subsection discusses two crucial aspects of PDCs, namely coverage and capacity.
The reviewed papers are summarized in Table~\ref{tab:CoverageCapacity}.

\subsubsection{Coverage}
{ In post-disaster scenarios, coverage can be compromised due to the failure of one or multiple cell towers.
Although several non-terrestrial technologies allow for restoring connectivity, the main challenge is timely putting them in operation.
Although these technologies can quickly cover large areas (thanks to the higher altitude compared to cell towers), their operation can be complex and last for long periods.
Nonetheless, their potential attracted academia and industry entities, leading to continuous technological improvements that partially solve the challenges mentioned above. }\par
D2D communication along with UAVs can be  used to provide ubiquitous coverage in post-disaster situations. Most of the existing works focus on using D2D for communicating in lack of a functioning infrastructure~\cite{Zhao19,marttin18energy,usman15D2D,christy17path,rawat15towards}.
For instance, authors in~\cite{marttin18energy} have proposed an LTE-based D2D technology and evaluated its performances for a typical disaster scenario. Authors in~\cite{Zhao19}, instead, have proposed establishing multihop D2D links to extend the coverage area of UAVs when there is a lack of functioning cell towers.
Furthermore, Ref.~\cite{usman15D2D} suggested a hierarchical D2D architecture with a centralized SDN controller communicating with the cloud head to minimize energy consumption. Also, authors in~\cite{christy17path} considered deploying UAVs to discover D2D devices in disaster-struck regions.
Finally, authors in~\cite{rawat15towards} provided an overview of the use of ubiquitous mobile devices and applications in the context of 5G for post-disaster communications.
Ref.~\cite{deepak19robust}, instead, proposed to combine D2D and cellular technologies to improve the coverage probability. 
The work proposed in~\cite{xu20time} introduced a performance metric called time-weighted coverage (since the behavior of the network is dynamic) to promote a novel 3D networking architecture with both terrestrial and aerial nodes.
The concept of coverage has also been discussed in~\cite{liu19transceiver} since UAVs are proposed as an effective solution to make up for the eventual loss of coverage for IoT applications.
Authors in~\cite{pfeiffenberger15coverage} studied via simulations the optimal deployment of wireless gateways and relay nodes while considering its influence on field commanders' positions. 
\subsubsection{Capacity}
Whenever there is a failure in the network infrastructure, not only coverage but also capacity becomes a critical issue.
{ In fact, cell towers cannot be entirely replaced by ad hoc nodes since the latter relies on wireless backhaul links and can usually carry a much smaller number of antennas while suffering delays due to their high altitude.
Moreover, current dedicated PPDR networks are usually based on terrestrial trunked radio (TETRA) and analog private mobile radio (PMR), which are more suitable for advanced voice services rather than data-intensive applications~\cite{casoni2015integration}.}
This is because the entire load has to be divided among only the surviving BSs, leading to outages and low data rates.
Also for this topic, very few relevant research articles have been published recently and are mentioned in what follows.
In~\cite{Matracia21disaster}, the deployment of the best aerial fleet in order to maximize the capacity in the middle of the disaster-struck area has been investigated.
Ref.~\cite{dai20constellation}, instead, included design considerations for a capacity-oriented LEO satellite constellation and implemented a multi-objective genetic algorithm that optimizes a combination of constellation cost, capacity, and multiple-coverage.

\begin{table}[]
    \centering
    \caption{Summary of the relevant papers on coverage and capacity}
    \begin{tabular}{|c|c|l|c|}
    \hline
    \textbf{Topic} & \textbf{Year} & \textbf{Main focus} & \textbf{Ref.} \\ \hline
    \multirow{9}{*}{Coverage} & \multirow{3}{*}{2015} &  \makecell[l]{Optimal deployment, wireless gateways,\\relay nodes} &~\cite{pfeiffenberger15coverage} \\ \cline{3-4} 
    & & D2D, SDR &~\cite{rawat15towards} \\ \cline{3-4}  
    & & D2D, SDN &~\cite{usman15D2D} \\ \cline{2-4} 
    & 2017 & D2D, UAVs &~\cite{christy17path}  \\ \cline{2-4} 
    & 2018 & D2D, LTE &~\cite{marttin18energy} \\ \cline{2-4} 
    & \multirow{3}{*}{2019} & D2D, multihop, UAVs &~\cite{Zhao19} \\ \cline{3-4}  
    & & D2D, 5G &~\cite{deepak19robust} \\ \cline{3-4}  
    & & UAVs, IoT &~\cite{liu19transceiver} \\ \cline{2-4}
    & 2020 & Time-weighted coverage, HetNets &~\cite{xu20time} \\ \hline
    \multirow{2}{*}{Capacity} & 2020 & Satellites, cost, multiple-coverage &~\cite{dai20constellation}  \\ \cline{2-4}
    & 2021 & VHetNets &~\cite{Matracia21disaster} \\ \hline
\end{tabular}
\label{tab:CoverageCapacity}
\end{table}

\subsection{Radio Resource Management} \label{subsec:radioRes}
{ Radio resource management can be a critical issue in the context of  PDCs. 
For example, in an aerial network deployment with FSO links, the performance can be significantly degraded in the case of polluted air (especially in the presence of smoke, fog, sand, or dust).
A potential solution to such a problem would be to introduce a complimentary wireless channel less affected by such conditions as RF transmission, as suggested by the authors of Ref. \cite{selim2018post}. Indeed, the paper highlighted that commercial hybrid FSO/RF systems already combined millimeter wave (mmWave) and laser-based FSO to serve large areas with at least one gigabit per second (Gbps) of data transmission. Hence, developing novel spectrum management techniques is crucial to overcoming the environmental effects in a post-disaster situation.} \par
For instance, ~\cite{deepak2019overview} discussed spectrum allocation strategies and future technologies for EMSs.
Ref.~\cite{liu19resource} introduced aerial network access and resource allocation scheme that optimizes the number of human portable/wearable machine-type devices (HMTDs) that are transmitting data. In the same context, authors in~\cite{nishiyama17OFDM} proposed a radio resource management system based on orthogonal frequency-division multiplexing (OFDM) using a novel resource allocation technique for improving aerial networks' data communications.
Moreover, in~\cite{carlberg19slicing}, a novel 5G network architecture based on slicing is introduced for supporting FRs' communications by dynamically prioritizing their channels, depending on their need.
The authors of Ref.~\cite{perez19supporting}, instead, investigated the use of radio access network (RAN) slicing and introduced management mechanisms that allow handling the slice reconfigurations.\par
For aerial networks, an autonomous spectrum management scheme based on spectrum sharing has been introduced and validated via simulation results in~\cite{shamsoshoara20an}.
Finally, authors in~\cite{nguyen19real} proposed a resource allocation scheme for UAV-enabled cellular networks as well as efficient algorithms for clustering selection and resource allocation. 


\subsection{Localization}
\begin{figure}
 \centering
    \includegraphics[width=0.6\columnwidth, trim={7.7cm 0.8cm 8.4cm 3cm}, clip]{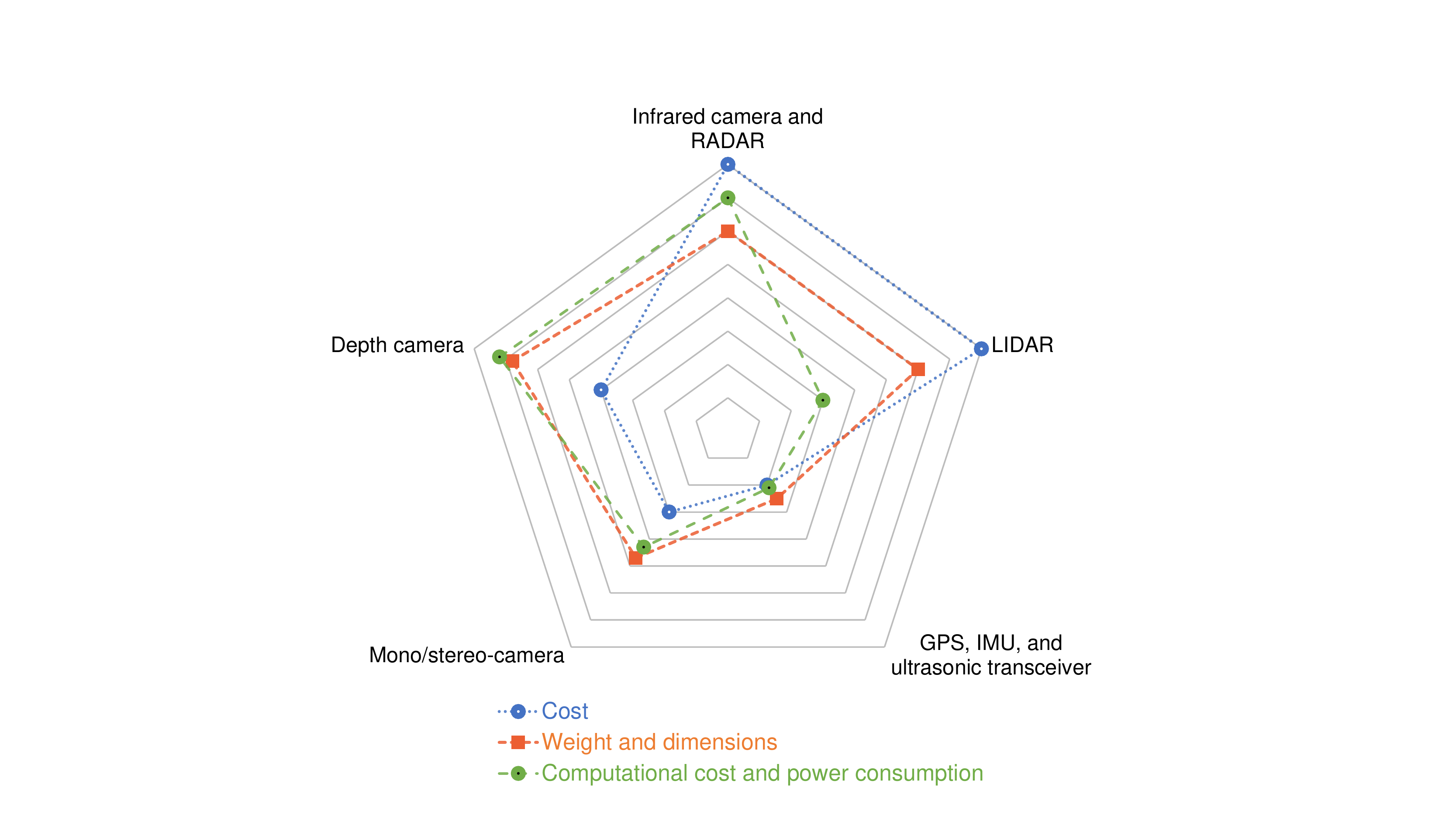}
    \caption{Qualitative comparison between the common SLAM sensors for UAVs \cite[Sec. 5]{recchiuto18assessment}. Note that infrared is hereby abbreviated as IR and RGB-D is a specific type of depth camera.}
    \label{fig:radar_sensors}
\end{figure}

Humans affected by a disaster must be rescued within the so-called golden 72 hours after the disaster. 
{ In this context, localization plays a major role in emergency scenarios, especially when a considerable percentage of the victims are trapped \cite{KHALIL2022}. It is evident that localizing the victims becomes much more challenging if they are trapped under rubble in case of earthquakes or covered by the smoke of a fire.
To further complicate the situation, disasters generally modify the environment, and thus existing maps might not be helpful anymore. This may require simultaneous localization and mapping (SLAM) techniques, which may overload the computational capability of the systems (\it{e.g.}, UAVs or other autonomous robots) and considerably increase their power consumption.
In order to perform SLAM tasks, UAVs should be equipped with sensors such as a global positioning system (GPS) receiver, inertial measurement unit (IMU), ultrasonic transceiver, light detection and ranging (LIDAR), radio detection and ranging (RADAR), and various types of cameras (\it{i.e.}, mono-, stereo-, infrared, and depth cameras). However, some of these sensors have strong techno-economic limitations, as summarized in Fig.~\ref{fig:radar_sensors}.
} \par
In general, localization techniques can be categorized as centralized (if the input data are processed by the BS) or decentralized (if the input data are processed by the sensor nodes)~\cite{Nasir2019}.
Due to the susceptibility of cell towers as well as the limited availability of power in disaster circumstances, our interest is mostly oriented towards decentralized localization techniques which can be further distinguished as range-based and range-free techniques.
Range based techniques require accurate measures of distances or angles between the devices of the network, and therefore they require an additional hardware as well as a stronger power supply when compared to range-free techniques~\cite{mehannaoui19study}, which however are usually less precise. 
Authors in~\cite{bhowmik16fuzzy} introduced a range-free scheme for localization in WSNs, which makes use of fuzzy logic to relate the received signal strength (RSS) and the distance so that the location can be evaluated in sufficiently precise manner. 
Following the same lines, the process of localization and the related procedures were presented in~\cite{mehannaoui19study} together with the taxonomy of range-free techniques, with a special focus on the so-called \it{DV-HOP} algorithm. 
Furthermore, Ref.~\cite{zrelli17localization} implemented two WSN-based methods to ensure the localization of vibration damage in tunnels. 
Finally, authors in~\cite{zuo20directional} focused on the estimation of both position and transmission orientation of a directional source in 3D WSNs. \par
In the specific context of PDCs, Ref.~\cite{gu13challenges} provided an insightful overview of the existing localization algorithms for post-disaster scenarios using WSNs. 
Authors in~\cite{ahmad15monitoring} proposed one possible approach for localizing damages and humans in a disaster-struck region.
Furthermore, Khan \it{et al}.~\cite{khan17location} devised a localization technique based on RSS measurements, where the relative ground coordinates of users are computed using \it{Isomap} (that is, a nonlinear technique applied for dimensionality reduction) and then transformed to the actual global coordinate system. \par
Alternatively, authors in~\cite{chai13localization} introduced a modeling and simulation method of the radio channel for rescue purposes, which can be used to develop radio localization systems. In general, the passive localization schemes that are developed in various works can be applied to post-disaster situations. 
For instance,~\cite{jahangiri20parametric} recently discussed localization in a WSN for different applications.
In the same context, authors in~\cite{zhang14node} implemented their algorithm by using PSO and promoted the path planning strategy based on a grid scan. 

\begin{table}[]
    \centering
    \caption{Summary of the relevant papers on localization}
    \begin{tabular}{|c|l|c|}
    \hline
    \textbf{Year} & \textbf{Main focus} & \textbf{Ref.} \\ \hline
    2013 & WSNs, algorithms &~\cite{gu13challenges} \\ \hline
    2014 & PSO, path planning strategy &~\cite{zhang14node} \\ \hline
    \multirow{2}{*}{2015} & Satellite stereo images &~\cite{ahmad15monitoring} \\ \cline{2-3}
    & Radio channel, rescue &~\cite{chai13localization} \\ \hline
    2016 & Fuzzy logic, range-free localization, WSNs &~\cite{bhowmik16fuzzy} \\ \hline
    2017 & Vibration damage, WSN &~\cite{zrelli17localization} \\ \hline
    2019 & Range-free localization, taxonomy &~\cite{mehannaoui19study} \\\hline
    \multirow{2}{*}{2020} & Gauss-Newton algorithm, accuracy, WSNs &~\cite{jahangiri20parametric} \\ \cline{2-3}
    & Directional source, 3D WSNs &~\cite{zuo20directional} \\ \hline
    \end{tabular}
    \end{table}

\subsection{Energy Efficiency} \label{subsec:en_eff}
Since the power infrastructure is also susceptible to calamities, emergency circumstances require energy efficient systems~\cite{udreakh18RAPID,li18UAV,do-duy21joint,nakamura19temporary,shah19bio,seng20solar,selim2018post,ashok20uninterrupted,rosas15mobility,li15renewable,kayiram21energy,ali18power,ngo14MDRU,mahmoud18topology}.
{ The nodes in a post-disaster network can rely on either renewable energy (especially photovoltaic panels), RF charging, batteries, or laser power beaming, to name a few.
However, these technologies are currently unable to ensure a continuous and prolonged service.
Nonetheless, many works have recently devised techniques that reduce the overall power consumption of PDCs systems, which can be summarized as follows.} \par 
Ref.~\cite{nakamura19temporary} proposed a disaster-time system that makes use of a message ferry method to collect and exchange information while improving DTNs' energy efficiency; in particular, the authors introduced a method that allows to relay the messages of mobile terminals with low battery level to the terminals with higher battery level, avoiding the former to fully discharge.
Inspired by biological networks of living organisms, authors in~\cite{shah19bio} introduced an energy-efficient disaster response network (DRN), called \it{Bio-DRN}, modeled via an integer linear programming optimization problem. In the same work, the Bio-DRN was developed by means of a sub-optimal heuristics and tested via simulation considering a real disaster-prone region in Nepal. { Another interesting work in~\cite{babu20energy}, optimized the 3-D placement of a set of aerial access points for energy-efficient uplink communications\footnote{The latter work is general and not limited to post-disaster scenarios.}.}\par
Authors in~\cite{seng20solar} investigated the possibility of mounting solar panels on top of UAVs to increase their autonomy while taking into account routing, data rate, and transmit power constraints. Evidently, the main issue with this solution is represented by the additional payload due to the panels themselves.
On the other hand, authors in~\cite{selim2018post} promoted a heterogeneous fleet that includes untethered drones for cellular coverage, tethered drones for high-capacity backhauling, and untethered powering drones for charging.
In~\cite{ashok20uninterrupted} a novel beamforming architecture, based on conditional time split-energy extraction (CT-EE) for enhancing nodes' autonomy, was presented and compared to conventional beamforming and other energy extraction methods. \par
Furthermore, to minimize both message overhead and energy consumption,~\cite{rosas15mobility} proposed a hybrid solution for DTNs where the routing protocol is chosen depending on the mobility patterns of each node.
Works such as~\cite{li15renewable}, instead, focused more on optimizing data traffic throughput. In particular, the authors described the problem via mixed integer linear programming (MILP) and developed a traffic demand-aware off-line energy-efficient scheme for WMNs constituted by renewable-energy-enabled base stations (REBSs).
Moreover, in the context of WSNs for disaster monitoring, Ref.~\cite{kayiram21energy} presented an energy-efficient data retrieval scheme based on intelligent sleep scheduling. In addition, the authors mathematically proved that the proposed scheme is capable to extend the longevity of the network, while contributing with traffic reduction and load balancing.
An interesting problem tackled with stochastic geometry is~\cite{ali18power}. 
Here, the authors used energy harvesting and transfer for the user equipment in D2D clustering communications for disaster management; the required power would be captured from RF signal via BS.
Similarly Ref.~\cite{saif21efficient} used the optimal cluster head (CH) technique to make energy transfer more efficient in UAV-assisted D2D PDCs.
The proposed results also showed improvements in terms of network coverage and reliability.
The novelty of the work presented in~\cite{ngo14MDRU}, instead, consists in combining both spectrum and energy efficiencies (thus, a new metric called spectrum-energy efficiency was defined) for renewable-energy-enabled gateways and MDRUs deployed in disaster struck environments. In particular, the authors proposed a topological scheme based on the top $k$ spectrum-efficient paths and showed how to optimize the value of $k$ itself.
Finally, authors in~\cite{mahmoud18topology} developed an algorithm for network reconfiguration in underwater communication systems that are capable of harvesting energy in case of disasters occurring in the ocean, such as tsunamis. 

    \begin{table}
    \centering
    \caption{Summary of the relevant papers on energy efficiency}
    \begin{tabular}{|c|l|c|}
    \hline
    \textbf{Year} & \textbf{Main focus} & \textbf{Ref.} \\ \hline
    2014 & MDRU, spectrum efficiency &~\cite{ngo14MDRU} \\ \hline
    \multirow{2}{*}{2015} & Mobility, DTN protocols &~\cite{rosas15mobility} \\ \cline{2-3}
    & Data traffic throughput &~\cite{li15renewable} \\ \hline
    \multirow{3}{*}{2018} & Energy harvesting, D2D &~\cite{ali18power} \\ \cline{2-3}
    & Energy harvesting, underwater IoT &~\cite{mahmoud18topology} \\ \cline{2-3}
    & Drones, battery, backhaul &~\cite{selim2018post} \\ \hline
    \multirow{2}{*}{2019} & Temporary communication system, DTN &~\cite{nakamura19temporary} \\ \cline{2-3}
    & Resilient networks &~\cite{shah19bio} \\ \hline
    \multirow{3}{*}{2020} & Optimal placement &~\cite{babu20energy} \\ \cline{2-3}
    & Solar powered UAVs &~\cite{seng20solar} \\ \cline{2-3}
    & Beamforming, MIMO &~\cite{ashok20uninterrupted} \\ \hline
    2021 & Data retrieval &~\cite{kayiram21energy} \\ \hline
    2021 & Energy harvesting, UAV, D2D &~\cite{saif21efficient} \\ \hline
    \end{tabular}
    \label{tab:energyefficiency}
\end{table}

\subsection{Important Remarks}
{ The presence of obstacles generated by the calamity (combined with the typical complexity of ad hoc network architectures) makes channel modeling extremely challenging. 
Due to this, accurate estimation of performance metrics such as joint access and backhaul coverage probability and capacity is not possible yet. 
This issue, together with the increased traffic demand and limited bandwidth and energy resources, strongly affects also radio resource management and energy efficiency (which is actually the most important requirement, since most devices are not designed to operate for a long time without any power infrastructure's energy supply).
}

\section{Networking Layer Issues} \label{sec:networkingLayer}
This section reviews various references on networking aspects including space-air-ground integration (from its architecture to its inherent complications), routing algorithms, and applications of DTNs and edge computing in emergency scenarios.

\subsection{Integrated Space-Air-Ground Architectures}
Space-air-ground integrated network (SAGIN) is an emerging paradigm that can be implemented also in post-disaster scenarios as a solution for improving security and extending coverage~\cite[Sec.~IX-B]{kodheli21satellite}.
The main idea behind this solution is to conveniently combine the three said layers: 
in fact, terrestrial networks have the lowest delays and the highest capacity (often without energy constraints), while satellites benefit from an extremely wide coverage area and they are generally resilient to any disaster; aerial platforms, instead, have intermediate characteristics between terrestrial and space nodes. 
For this paradigm, the space, air, and ground segments can either inter-operate or be independent of each other.
{ If they inter-operate, one of the main challenges (essentially due to the different trajectories and velocities between the nodes) consists in effectively harmonizing the flight of the aerial platforms (especially if both HAPs and LAPs are present) so that the risk of getting them disconnected is minimized: to achieve this, novel routing protocols taking into account the variety of the interconnections (\it{e.g.}, hybrid RF/free space optics between HAPs and LAPs, and free-space optics among HAPs) need to be designed~\cite[Sec.~IX-B]{kodheli21satellite}.}\par
One typical example of SAGIN architecture is the so-called \it{Global Information Grid (GIG)}, which is made by four layers (\it{i.e.}, ground, aerospace, near-space and satellite layers) embedding the communication, sensor, and operation networks~\cite{liu18space_survey}.
Generally speaking, the space network may include geosynchronous equatorial orbit (GEO), MEO, and LEO satellites, with the respective terrestrial infrastructures (\it{e.g.}, ground stations and control centers).
Note also that the terrestrial layer is not limited to cell towers, but may also include MANETs, wireless local area networks (WLANs), \it{etc.} \par
In Ref.~\cite{akiyoshi17STICS}, a novel cooperative communication scheme for UAV-aided satellite/terrestrial integrated mobile communication systems (STICSs) is proposed as a means of interference mitigation.
Indeed, the UAVs act as relay stations and, according to the simulation results obtained, allow to reduce the average bit error rate (BER) and boost the throughput of the system.
Authors in~\cite{pirzada20detection}, on the other side, provided a framework for using Internet Protocol (IP) for disaster management services within a SAGIN. 
The study simulated two disasters occurring in Africa and North America, and showcased the additional services offered by the proposed IP-based method.

\subsection{Routing}
{ Generally speaking, the problem of routing consists of finding the optimal path that effectively transfers data among the network nodes. It is evident that post-disaster wireless environments are incredibly unpredictable, and therefore routing cannot be planned as carefully as in ordinary circumstances.
As far as we are concerned, the main routing challenges brought by the occurrence of a calamity are due to terrestrial nodes' malfunction or failure, inaccurate placement of ad-hoc nodes, unpredictable traffic demand and distribution, energy shortage, and harsh environmental conditions (which may essentially affect the wireless communication channels and, for example, reduce the ad-hoc nodes' coverage radii).
}
Since routing can tremendously impact on the performances of the network (especially in terms of energy efficiency and delay), several works have recently tackled this topic, as discussed in the rest of this section. \par
Authors in~\cite{chaudri19green} have improved the multicast routing squirrel search algorithm (SSA) for providing green communications and has experimentally shown the effectiveness of properly balancing energy consumption and other quality of service (QoS) parameters.
The study proposed in~\cite{Xu20fast}, instead, relied on real-world maps and suggested several methods on access point placement and routing in order to quickly connect users inside middle-size disasters.
Moreover, authors in~\cite{Masaracchia19framework} proposed a D2D-based framework to cluster users' devices and optimize the transmission power for each gateway. 
One of the most remarkable strategies consists in putting the nodes within the damaged area in \it{LISTEN} mode and provide them the clustering instructions from a functional area, thus saving valuable energy to the devices. \par
In the context of dual-channel-based MANETs, authors in~\cite{kim20dual} recently introduced an algorithm for efficient routing, since each node is able to configure the routing table based on the exchanged neighbor list.
The main application suggested for this work is indoor communications for firefighters, especially since previous works do not consider the mobility of the nodes properly, and neglect the presence of potential obstacles in the environment.
Another approach based on routing table has been proposed in~\cite{uddin13intercontact}, where the concept of \it{intercontact routing} was introduced to estimate delivery probabilities and route delays, as well as to find reliable routes and consequently control message replication and forwarding.
Furthermore, the authors enhanced the energy efficiency of protocol by means of a differentiated message delivery service. \par
For UAV networks, instead, Ref.~\cite{arafat18locationAided} suggested a location-aided delay-tolerant routing (LADTR) protocol in order to exploit both store-carry-forward (SCF) technique and location-aided forwarding.
Indeed, this work improved the efficiency of SCF by introducing ferrying UAVs into the network, and validated the LADTR protocol by numerically comparing it (in terms of routing overhead, packet delivery ratio, and average delay) with other common protocols. \par
Authors in~\cite{uchida17triage} focused on DTNs and introduced additional routing methods (\it{e.g.}, \it{Node Selection} by the evacuator's territory, \it{Data Triage} by data priority, and \it{Dynamic FEC} controls by the Jolly-Sobor model).
In this work, field experiments were carried out for validating the performances of the proposed routing methods.
Morover, the authors of Ref.~\cite{rosas15mobility} suggested hybrid DTN protocols for allowing nodes to apply different routing rules depending on their own mobility patterns. 
Instead, authors in~\cite{kang18ICN} introduced and analyzed a DTN routing protocol for information-centric networks (ICNs) in disaster-struck areas, showing its advantages in terms of delivery probability and overhead ratio.
Finally, the D2D-based architecture proposed in~\cite{usman15D2D} uses a SDN controller which can conveniently enable multi-hop routing path between victims and FRs.\par
To deepen this topic, the reader could also refer to \cite{Jahir19routing}, which surveyed routing protocols for MANET-based architectures in disaster area networks (DANs). 

\begin{table} []
    \centering
    \caption{Summary of the relevant papers on routing}
    \begin{tabular}{|c|l|c|}
    \hline
    \textbf{Year} & \textbf{Main focus} & \textbf{Ref.} \\ \hline
    2013 & Intercontact delay &~\cite{uddin13intercontact} \\ \hline
    \multirow{2}{*}{2015} & D2D, SDN &~\cite{usman15D2D} \\ \cline{2-3}
    & Hybrid DTN protocols, mobility &~\cite{rosas15mobility} \\ \hline
    2017 & Experiments on routing methods &~\cite{uchida17triage} \\ \hline
    \multirow{2}{*}{2018} & Routing protocol for UAV networks &~\cite{arafat18locationAided} \\ \cline{2-3}
    & DTN routing protocol for ICNs &~\cite{kang18ICN} \\ \hline
    \multirow{2}{*}{2019} & SSA &~\cite{chaudri19green} \\ \cline{2-3}
    & D2D transmit power optimization &~\cite{Masaracchia19framework} \\ \hline
    \multirow{2}{*}{2020} & Access point placement &~\cite{Xu20fast} \\ \cline{2-3}
    & Dual-channel-based MANETs &~\cite{kim20dual} \\ \hline
    \end{tabular}
    \label{tab:routing}
\end{table}

\subsection{DTNs}
Whenever there is no way to ensure a reliable backhaul link to the core network, connectivity is compromised.
Since this situation is recurrent in emergency scenarios, DTNs can play a crucial role in the context of disaster management.
Therefore, many researchers started investigating the potential applications of DTNs in this area.
{ However, it is evident that DTNs rarely operate in real time, therefore devising novel and powerful routing algorithms is needed in order to ensure a sufficient QoS (while meeting the energy and memory constraints of IoT devices) in emergency situations.
Moreover, routing protocols should use context information about traffic to influence the relay selection depending on the user traffic needs~\cite{krug16challenges}.} \par
In~\cite{fauzan19IBR}, the so-called \it{IBR-DTN} architecture (\it{i.e.}, an implementation of the bundle protocol RFC5050) has been shown to be effective when sending small size data such as text messages.
On the other side, if the connection is lost during the transmission then the entire file would need to be sent again, which makes this architecture very inefficient for sending heavy files.
Ref.~\cite{ganguly15location} has suggested to use delay-tolerant networking in a location-based mobility prediction scheme that estimates the mobility pattern of the nodes (\it{e.g.}, FRs or victims equipped with smart devices) and enables to select the best forwarder.
Another interesting work has been presented in~\cite{roy16health}, where the authors promoted an infrastructure-less health data delivery process architecture capable of automatically identify injured persons.
In order to make up for the eventual unavailability of cloud-based mapping services and data in post-disaster scenarios, authors in~\cite{trono15MapEx} have presented \it{DTN MapEx}: a distributed computing system that generates and shares maps over a DTN.
To do this, users need to log the GPS traces of their routes and collect data about the disaster-struck environment; then, pre-deployed computing nodes process the collected data to generate a map for the network.
Ref.~\cite{kawamoto15message} devised a novel DTN-based message relay protocol that incorporates message delivery into a specific type of shelter network called \it{autonomous wireless network construction package with intelligence (ANPI)}.
This solution was tested via simulations, showing its effectiveness in reducing redundant transmissions and improving the message delivery ratio. \par
The authors of~\cite{basu19principal} derived a principal component regression model and proposed an opportunistic demand sharing scheme for collecting and spreading resource demands to the control station via a smartphone-enabled DTN.
By means of selected case studies, this work also illustrated to what extent DTNs can be useful for demand forecasting.
More recently, the concept of DTN has been applied also for post-disaster resource allocation, used in a novel opportunistic knowledge sharing approach for gathering the resource needs in a utility-driven system~\cite{basu20utility}; simulations results showed that the proposed system is very competitive with the similar ones proposed in literature, especially in a fully connected scenario. 
Ref.~\cite{krug16challenges} used realistic traffic and mobility data in order to evaluate various routing schemes for DTNs, and introduced the option of combining dedicated DTN routing to additional aerial nodes.
Finally, authors in~\cite{hoque20SDN_DTN} proposed a four-layer architecture consisting of a combination of SDN and DTN in order to minimize packet loss.
Interestingly, the \it{Mininet-WiFi} simulation results suggested that the proposed architecture can achieve a packet loss as low as $0.46\%$.\par
A quick summary of the aforementioned papers is proposed in Table~\ref{tab:DTN}.

\begin{table} []
    \centering
    \caption{Summary of the relevant papers on DTNs}
    \begin{tabular}{|c|c|l|c|}
    \hline
    \textbf{Topic} & \textbf{Year} & \textbf{Main focus} & \textbf{Ref.} \\ \hline
    \multirow{9}{*}{DTNs} & \multirow{3}{*}{2015} & Mobility prediction &~\cite{ganguly15location} \\\cline{3-4}
    & & Mapping &~\cite{trono15MapEx} \\ \cline{3-4}
    & & Message delivery ratio &~\cite{kawamoto15message} \\ \cline{2-4}
    & \multirow{2}{*}{2016} & Mobility &~\cite{krug16challenges} \\ \cline{3-4}
    & & Healthcare, identification &~\cite{roy16health} \\ \cline{2-4}
    & \multirow{2}{*}{2019} & Demand forecasting &~\cite{basu19principal} \\ \cline{3-4}
    & & IBR-DTN &~\cite{fauzan19IBR} \\ \cline{2-4}
    & \multirow{2}{*}{2020} & Resource allocation &~\cite{basu20utility} \\ \cline{3-4}
    & & \makecell[l]{Hybrid SDN-DTN,\\message delivery rate, packet loss} &~\cite{hoque20SDN_DTN} \\ \hline
    \end{tabular}
    \label{tab:DTN}
\end{table}

 \subsection{Edge Computing}
By definition, edge computing is an autonomous computing model that comprises many distributed devices communicating with the network for several computing tasks~\cite{vaquero14finding}. 
In other words, edge computing is a paradigm meant to move computational data, applications, and services from the cloud servers to the edge network to minimize latency and maximize bandwidth~\cite{khan19edge}. \par
{ Apart for the evident need of low latency communications we recall that the problem of bandwidth is also due to the susceptibility of FSO to harsh atmospheric conditions, as already mentioned in Sec.~\ref{subsec:radioRes}.}
{ Furthermore, UAV-enabled multi-access edge computing could provide additional resources and effective low-latency services compared to traditional MEC architectures (since UAVs can be easily moved closer to the sources of data generation).
However, designing efficient offloading and resource allocation strategies is difficult because of the heterogeneous protocols and interfaces used in this context~\cite{fatima22integration}.
In line with these goals, the National Science Foundation (NSF) announced the project \it{AirEdge}.
The main idea is to use a swarm of UAVs to carry mobile radio access points and edge servers, enabling on-demand edge computing and networking services for anti-terrorism, disaster rescue, or public safety applications~\cite{NSF20AirEdge}.}\par
Ref.~\cite{liu19edge} devised \it{Echo}, an interesting edge-enabled framework for disaster rescue, relying on computer vision to analyze and filter crowdsourced pictures to provide only relevant content to FRs and preserve bandwidth.
The authors also designed an adaptive photo detector to improve the precision and recall rate.
On the other hand,  authors in~\cite{xu20bigData} recently implemented mobile edge computing (MEC) task management strategies by applying long-range wide-area networking (LoRaWAN) to UAV-aided architectures.
The presented simulation results promoted this solution since it conveniently enables long-range MEC service. 
{ Another interesting work has been presented in~\cite{kaleem19uav}, where several PS-LTE enabling services were discussed in the context of the current \it{3rd Generation Partnership Project (3GPP)} standard releases.
Moreover, the authors proposed and evaluated a disaster-resilient edge architecture consisting of: SDN layer (to enable centralized control), UAV cloudlet layer (to perform either edge computing or enable emergency communications), and radio access layer.
{ Another interesting paper is~\cite{du21drone}, where a novel framework for drone-assisted and blockchain-enabled edge-cloud computing networks (DBECNs) is presented.
The authors devised an architecture (consisting of the terminal, multifunctional drone, blockchain, edge, and cloud layers) where UAVs act as moving terminals or BSs and proposed various use cases, including disaster relief and crowd monitoring.
However, important challenges related to AI-based algorithms, scalability and QoS, channel modeling, as well as the integration of SAGINs and edge-computing-enabled networks are faced by the proposed framework.}
}\par
Authors in~\cite{zhang21blockchain} proposed a data backup scheme to combine edge computing and blockchain technologies for disaster scenarios: while the former processes big data coming from the microgrid, blockchain provides security to the equipment needed to perform edge computing. However, the main challenge here is to deal with the trade-off between energy efficiency and security.
Moreover, authors in ~\cite{tran20named} designed and implemented a named-data-networking-based support system over the edge computing platform \it{KubeEdge}.
By doing this, this work showed the effectiveness of the proposed solution in promoting efficient emergency communications and responders' mobility.
Finally, authors in~\cite{hussain19federated} studied resource allocation to enable latency-intolerant tasks due to emergencies in oil fields. In particular, they devised a stochastic model that captures end-to-end uncertainties within the proposed federated edge environment to estimate the risk associated with each task and showed a performance improvement of almost 30 percent compared to state-of-the-art solutions.


\subsection{Important Remarks}
{ Generally speaking, heterogeneity brings strengths and challenges at the same time. 
Integrating different communication technologies and protocols in the same system requires innovative research because each node type is designed to operate in specific environments and conditions and may not be compatible with different types~\cite[Sec. 7]{khan22emerging}.
Also, networking issues might arise even in the case of simple multi-layered architectures, where any failed gateway node must be rapidly detected and properly replaced by another node for timely recovery~\cite[Sec. 4]{chen20review}.
Nonetheless, we believe that SAGIN architectures represent the most promising PDCs paradigm compared to DTNs because
they require more flexibility, which might be incompatible with the tight time constraints characterizing emergencies. 
Besides, UAV-enabled edge computing projects such as AirEdge could develop powerful technologies and expand networks' resources. Additionally, due to PDC networks' ad hoc structure, robust routing protocols need to be devised.
Also, we believe that also security plays an important role, especially in the context of human-made disasters (\it{e.g.}, terrorism and wars), and therefore should not be overlooked.
}

\section{Proposed Use Cases}
\begin{table} []
    \centering
    \caption{Main simulation parameters}
    \begin{tabular}{|l|l|}
    \hline
    \textbf{Description} & \textbf{Value} \\ \hline
    Number of system realizations & $2\!\times\!10^4$ \\\hline
    Simulation radius & $r_s=\begin{cases}
       r_d\!+\!3 \normalfont\text{ [km], for TBSs}\\
       h_S\, \normalfont\text{, for satellites}\\
       \end{cases}$ \\\hline
    Type of environment & Urban~\cite{al2014optimal} \\ \hline
    TBS density & $\lambda_T=10$ TBSs/km$^2$ \\\hline
    Path loss exponents &    $\begin{cases}
       \alpha_M=3 \\
       \alpha_T=2.9 \\
       \alpha_L=[2.5, 3]\blank\normalfont\text{, for [Los, NLoS]} \\
       \alpha_H=[2.2, 3]\blank\normalfont\text{, for [Los, NLoS]} \\
       \alpha_S = 2 
       \end{cases}$ \\  \hline
    LAP altitude &    
       $h_L=0.2 \normalfont\text{ km}$ \\  \hline
    Transmit powers &    $\begin{cases}
       p_T=p_M=10 \normalfont\text{ W}\\
       p_L=3 \normalfont\text{ W}\\
       p_H=20 \normalfont\text{ W}\\
       p_S = 1000 \normalfont\text{ W}
       \end{cases}$ \\  \hline
    Nakagami-m shape parameters & $\begin{cases}
       m_T=m_M=1\\
       m_L=2 \\
       m_H=m_S=3
       \end{cases}$ \\\hline
    SINR thresholds &  $\tau=\begin{cases}
       0.1\blank\normalfont\text{, for access links} \\ 0.2\blank\normalfont\text{, for backhaul links}
       \end{cases}$ \\  \hline
    Noise power & $\sigma_n^2=10^{-12}\,\normalfont\text{ W}$ \\\hline
    \end{tabular}
    \label{tab:parameters}
\end{table}

In this section, we present two use cases referring to the problem of coverage in the event of a disaster of either small or large scale.
{The main goal is to identify the optimal ad hoc network architecture (made of terrestrial, aerial, and space nodes) in order to serve a multitude of users inside the suffered region.}
Then, we extract valuable insights based on the proposed simulation results. \par
Both the considered setups (each one referring to a specific use case) assume uniformly distributed users and TBSs;
however, inside the disaster area $\mathscr{A}_0$ (assumed circular with radius $r_d$), the original infrastructure is nonfunctional, while outside it operates appropriately with a reliable connection to the core network (\it{e.g.}, by means of optical fiber links).
All the other backhaul links (except the one connecting the satellite to the ground station) are not ideal and thus cannot guarantee reliable communications. 
The QoS is evaluated based on averaged coverage probability experienced by the users inside the suffered region. 
A summary of the main system parameters is presented in Table \ref{tab:parameters}, where $h$ denotes the transceiver altitude and the subscripts $L$, $H$, $M$, $T$, and $S$ respectively refer to LAPs, HAPs, MDRUs, TBSs, and LEO satellites.

\subsubsection{Small Disasters}

\begin{figure*}
 \centering
    \includegraphics[width=1\textwidth, trim={0cm 0cm 0cm 3.5cm}, clip]{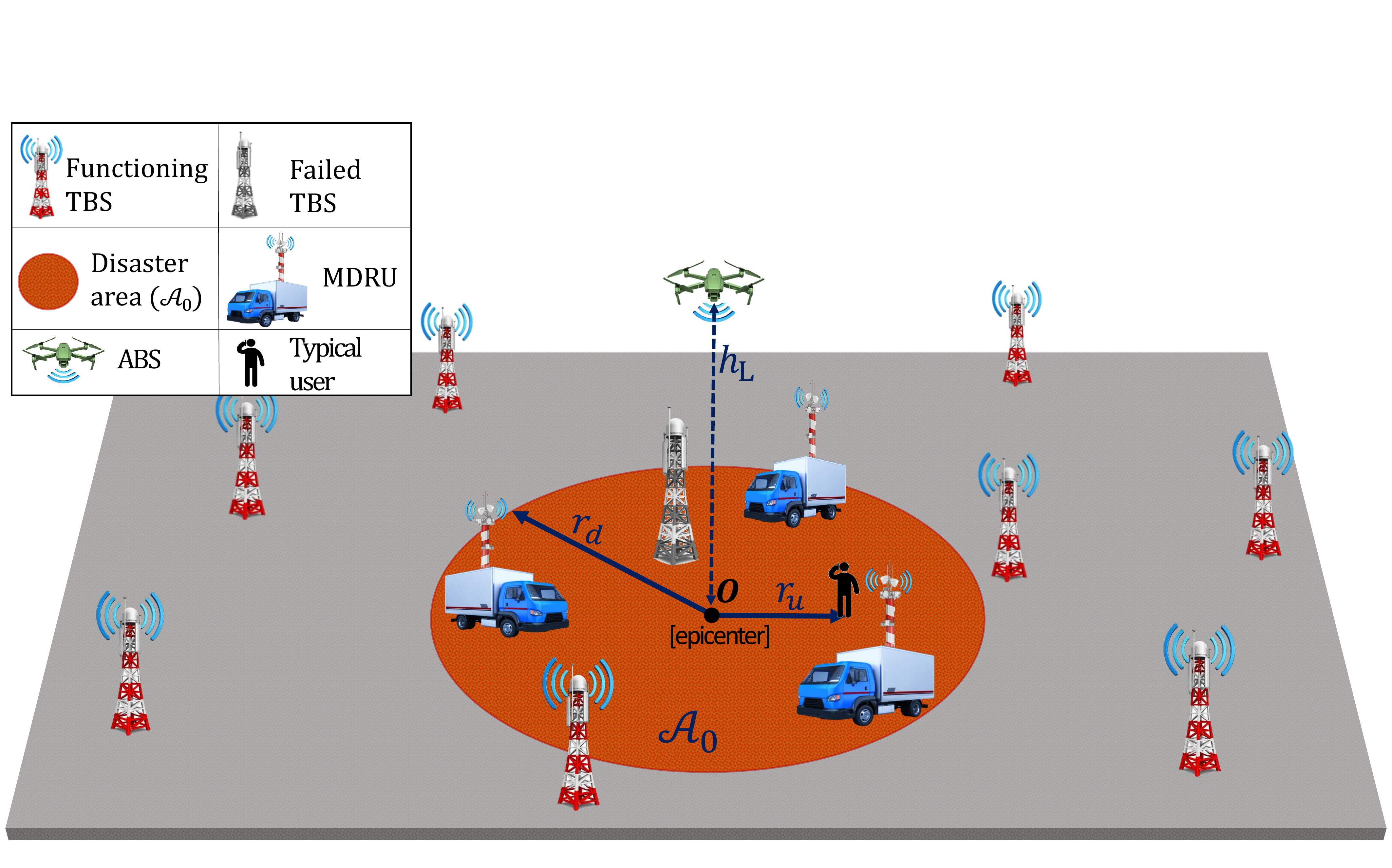}
    \caption{Proposed system setup for small disasters: one LAP hovering above the origin $O$ and a given number of MDRUs ($n_M$) are deployed in order to strengthen the disaster-struck network infrastructure.}
    \label{fig:setup1}
\end{figure*}

\begin{figure}
 \centering
    \includegraphics[width=0.65\columnwidth, trim={0cm 0cm 0cm 0cm}, clip]{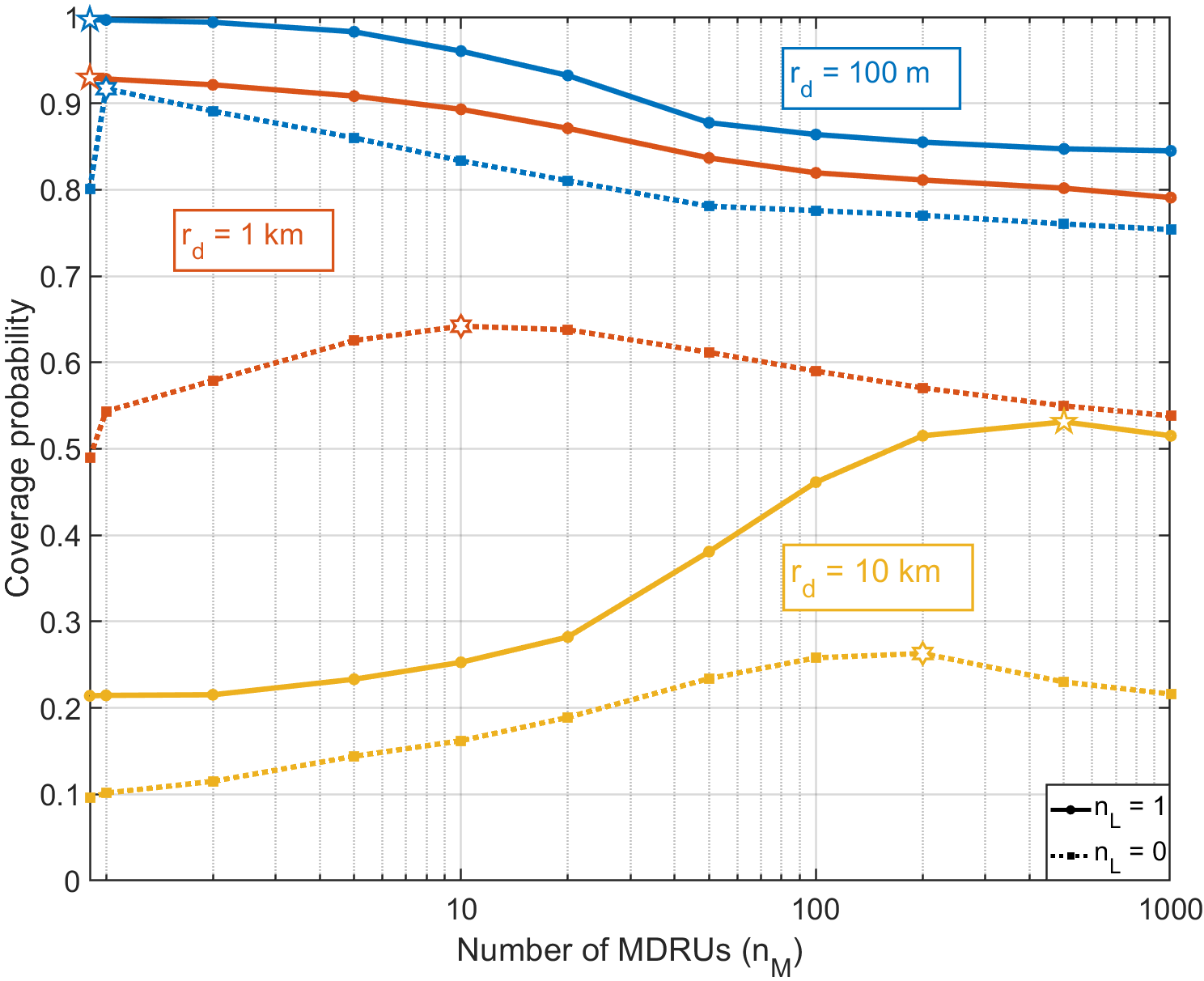}
    \caption{Simulated coverage probability for the first setup.
    The markers highlight the maximum coverage probability for each line.}
    \label{fig:simulation1}
\end{figure}

In this scenario, we assume that $r_d$ is relatively small (\it{i.e.}, from a hundred meters to ten kilometers, as in most cases), and thus LAPs are generally preferred over HAPs~\cite{Matracia21disaster}.
As depicted in Fig.~\ref{fig:setup1}, we propose a simple strategy that consists in deploying one single LAP above the disaster epicenter as well as a set of MDRUs (uniformly distributed over the disaster area and fixed in number). \par
{ Therefore, the users can associate with any of the three types of BS (outer TBS, LAP, or MDRU), but only TBSs have a direct connection to the core network.}
Hence, for each user served by a LAP or an MDRU at least one additional non-ideal wireless backhaul link is needed, which implies a higher risk of outage.
Based on the maximum average received power association rule for each consecutive link and assuming MDRUs to have only access functionalities, the possible paths are user--TBS, user--LAP, user--LAP--TBS, user--MDRU--TBS, and user--MDRU--LAP--TBS.
\par
The results displayed in Fig.~\ref{fig:simulation1} clearly show that the coverage probability strongly depends on $r_d\,$, and in particular, a larger disaster radius generally implies a lower QoS.
Indeed, for large values of $r_d$ the system suffers the fact that, given that the distance to the closest TBS often is larger than the respective coverage radius, one single drone is not sufficient to provide reliable backhaul for hundreds of MDRUs.
Therefore, the backhaul link becomes the bottleneck of the system. \par
Moving to more specific considerations, the simulated curves define two different cases:\\
\it{(i)} As long as $r_d$ does not exceed one kilometer (see the blue and the red lines), MDRUs are unnecessary since the user can easily associate to either the LAP or the closest TBS;\\
\it{(ii)} For much larger disasters (see the yellow lines), the optimal number of MDRUs ($n_M^*$) jumps to several hundreds, since average distance between the user and either the LAP or the closest surviving TBS becomes excessive; \\
\it{(iii)} The value of $n_M^*$ always increases as $r_d$ increases.
Moreover, if $r_d$ is sufficiently small, the presence of a LAP reduces $n_M^*$ since a considerable percentage of users directly associates to the aerial node and does not need further MDRUs; 
on the other hand, for $r_d=10\,$km the backhaul link offered by the LAP becomes vital and promotes the deployment of further ad hoc nodes, which otherwise would mostly backfire because of their limitations in reaching the core network.

\subsubsection{Large Disasters} \label{subsubsec:large_disast}

\begin{figure*}
 \centering
    \includegraphics[width=1\textwidth]{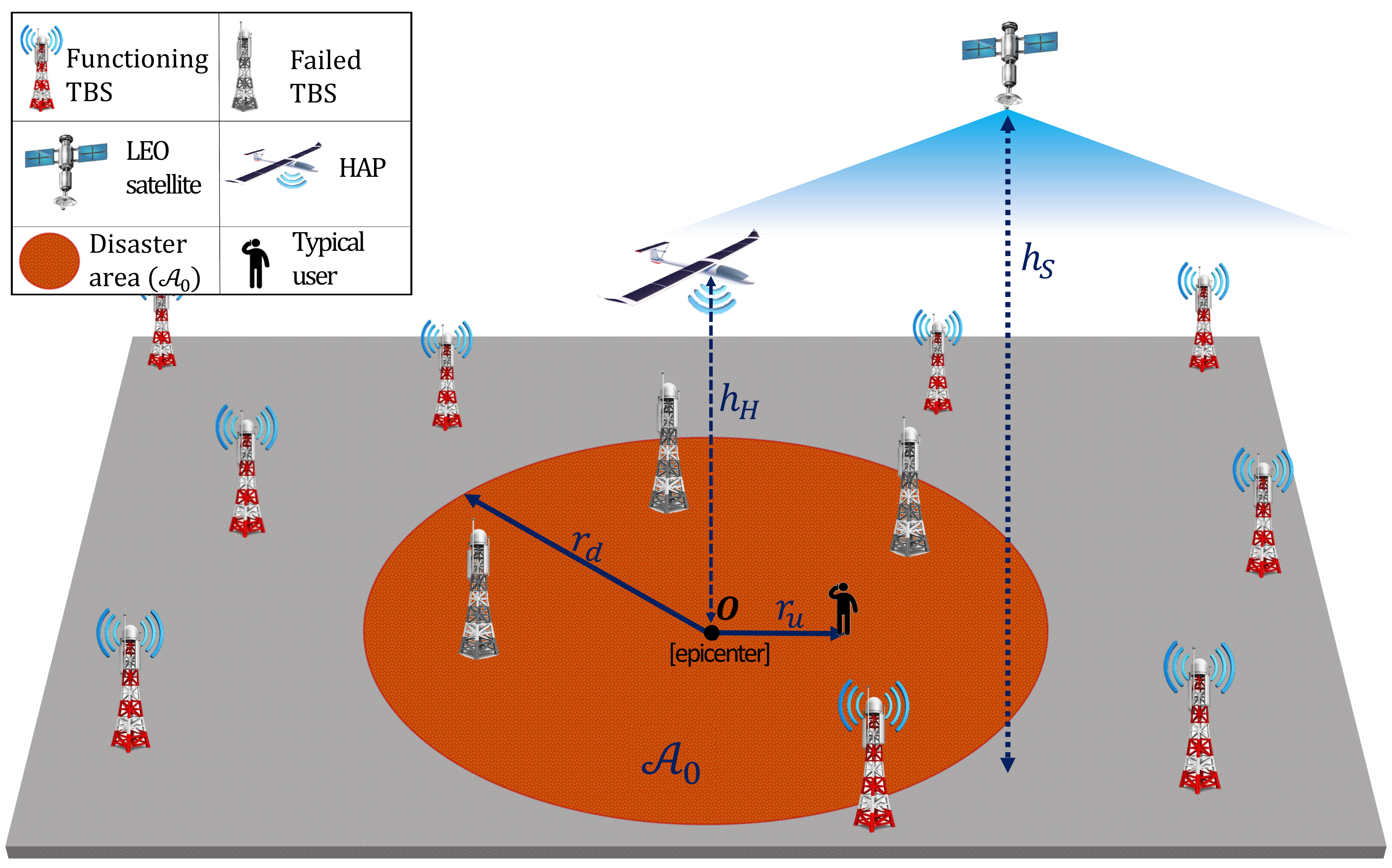}
    \caption{Proposed system setup for large disasters: one HAP hovering above the origin $O$ and one satellite around the area of interest support the disaster-struck terrestrial infrastructure.}
    \label{fig:setup2}
\end{figure*}

\begin{figure}
 \centering
    \includegraphics[width=0.65\columnwidth, trim={0cm 0cm 0cm 0cm}, clip]{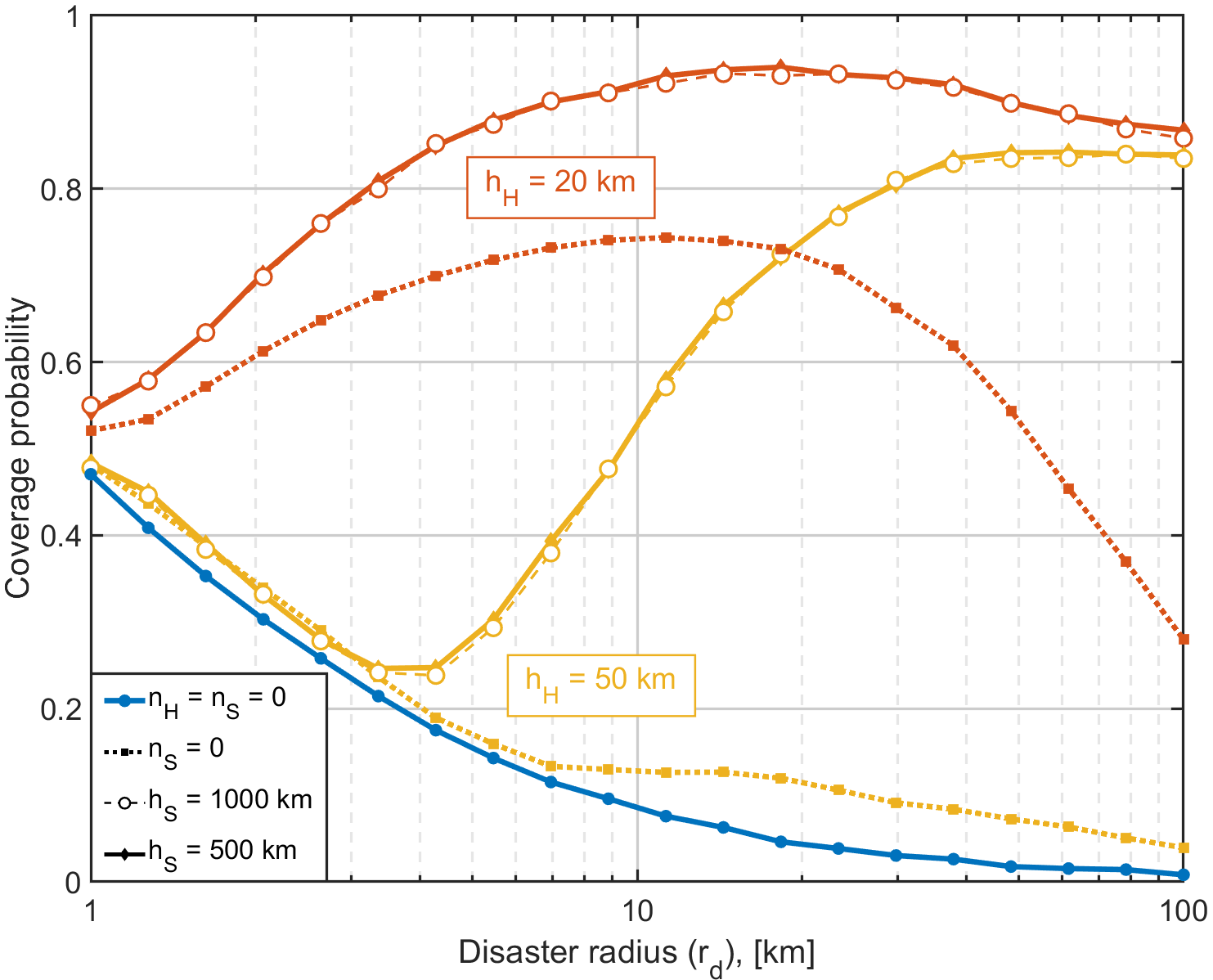}
    \caption{Simulated coverage probability for the second setup ($n$ denotes the number of nodes).}
    \label{fig:simulation2}
\end{figure}

The simulation results regarding first setup have evidently shown the limitations of the proposed strategy when $r_d$ exceeds a few kilometers.
Therefore, we hereby test a different network architecture that supports the surviving infrastructure by means of a HAP above the disaster epicenter and a LEO satellite (see Fig.~\ref{fig:setup2}). 
For the HAP-satellite link, the shadowed Rician fading channel is assumed according to~\cite{talgat21satellite}.\par
For this setup, we recall that the satellite backhaul connection to the closest ground station is assumed ideal.
However, the user can exploit the satellite only for backhaul, by associating to a HAP.
Therefore, the set of possible communication paths consists of just three elements: user--TBS, user--HAP--TBS, and user--HAP--satellite.
\par
We can note that Fig.~\ref{fig:simulation2} is consistent with Fig.~\ref{fig:simulation1}, meaning that, for $r_d=[1, 10]\,$km, the blue curve in the former matches with the initial values of the respective dotted lines in the latter.
Moreover, several insights can be extracted: \\
\it{(i)} It is evident that the satellite altitude $h_S$ does not have a considerable influence on the coverage probability (the yellow and red markers match with the respective solid lines). 
This suggests that, for any value of $r_d\,$, the air-to-space backhaul link is not the bottleneck of the system; \\
\it{(ii)} Despite its altitude is not relevant, the satellite starts playing a vital role as $r_d$ exceeds a dozen kilometers; \\
\it{(iii)} For $r_d\to1\,$km, the proposed scheme cannot provide a strong support to the terrestrial infrastructure (\it{i.e.}, the coverage probability falls between $40\%$ and $60\%$, no matter which of the proposed strategies is implemented);\\
\it{(iv)} For any value of $r_d\,$, the highest QoS can be achieved when a satellite is available and the HAP is deployed at relatively low altitude, despite the lower LoS probability.

\section{Challenges and Research Directions} \label{sec:Challenges}
Despite the research efforts discussed in the previous sections, there are still many open problems to address.
In particular, this section focuses on modulation and coding schemes, backhauling, optimal placement and trajectory/scheduling, and HO issues while suggesting possible future research directions.

\subsection{Modulation and Coding Schemes}
Another crucial aspect for studying the physical layer in PDCs regards the modulation and coding schemes to use.
Unfortunately, this area has not been strongly tackled yet\footnote{$\,$Because of this, we include this topic in the list of challenges for PDCs.}, but nonetheless insightful considerations have been presented for general applications. 
{ In PDCs, reliability should be a priority in emergency scenarios. 
Therefore, we believe that modulation and coding schemes that, given the bandwidth and energy constraints, ensure a low BER should be preferred over those that try to optimize performance metrics such as rate, latency, and throughput.
For example, let us consider the waveform orthogonal frequency division multiplexing (OFDM): 
schemes based on quadrature amplitude modulation (QAM) would be preferred over the ones based on phase-shift keying (PSK) because of the larger distances among their adjacent constellation points (although PSK can be more energy efficient). Moreover, it is worth recalling the importance of the channel type: for example, compared to the additive white Gaussian noise (AWGN) channel, in a Rayleigh channel, the BER is higher and less dependent on the SNR \cite{haboobi19modulation}.}\par
In the context of PDCs, the experimental results proposed in \cite{kojima14prevention} showed that siren sounds can be used in conjunction to complete complementary codes (CCCs) to embed messages that can be visualized by means of an \it{Android} application.
In this way, the information dissemination would be more effective (especially for hearing-impaired people), but nonetheless powerful error-correcting codes are needed whenever the bit rate exceeds roughly one thousand bits per second.
Authors in.~\cite{esposito20reinforced} presented a secure and inexpensive gossiping method enabled by reinforcement learning and game theory to mitigate the communication losses usually occurring in disaster scenarios.
Simulation experiments validated the benefits of this approach, compared to other schemes from the literature, in terms of overhead, latency, reliability, security, 
and energy efficiency.
Furthermore,~\cite{Surampudi2018a} presented the derivation of the mean co-channel interference for one-dimensional and bidimensional \it{LiFi balloon networks (LiBNets)} that are distributed according to a homogeneous Poisson point process (HPPP). 
Finally, the work presented in~\cite{shin17mixed} is a study about the mixed support of time division multiple access (TDMA) and single channel per carrier (SCPC) transmission to improve the bandwidth utilization in satellite-aided disaster communications.
The authors considered four different scenarios in order to categorize the formation of bandwidth assignment depending on the network's characteristics.

\subsection{Backhauling}
One crucial functionality of cell towers is providing backhaul connections to mobile users. 
Most of the ad-hoc solutions discussed above provide access only, meaning they still have to rely on TBSs or satellites for backhaul connectivity. 
{ However, it is impossible to equip the small devices (including drones~\cite{selim2018post}) with satellite transceiver equipment: to this extent, we believe that, whenever their implementation is feasible, tethered backhaul UAVs~\cite{Kishk20magazine} generally represent an excellent solution to the problem. 
Apart from intelligent allocation of resources such as bandwidth and energy~\cite{tezergil21wireless}, backhaul links also require the proper deployment of the network nodes to be reliable and effective.  
Usually, heterogeneous networks have higher complexity; meshed networks might be more convenient in emergency scenarios due to their higher level of redundancy and versatility. However, the inherent challenge in deploying large-scale self-organizing networks with UAVs is that the topology is subject to continuous modifications. Thus, fast computational methods are needed for PDCs management.}
To this extent, novel solutions are needed to make up for the potential failure of TBSs by ensuring reliable backhaul links for any other node of the network. \par
Motivated by the frequent catastrophes that occurred in the Philippines, Ref.~\cite{marciano16roger} devised a solution to provide GSM access and backhaul in the aftermath of a disaster. The idea is essentially to rapidly deploy a mini-cell tower and make it operate in IP protocol (\it{e.g.}, via TVWS or WiFi backhaul links) to serve FRs and victims.
Then,~\cite{alHilfi18wimax} suggested using worldwide interoperability for microwave access (WiMAX) technology to serve as backhaul to the WiFi network in disaster recovery circumstances by integrating it with WLAN access points.
In addition, the authors implemented voice over IP (VoIP) services in order to assess the system's capabilities. \par
Authors in~\cite{Kouzayha20hybrid} recently borrowed tools from stochastic geometry to estimate the effectiveness of mmWave backhauling for ABS-aided terrestrial networks. 
In particular, an algorithm was developed to maximize the throughput based on user association, UAV positioning, and resource allocation.
Furthermore,~\cite{yuan20joint} jointly optimized resource allocation, user association, and UAV positioning while taking into account both access and backhaul links, since the transmission rate of the system is given by the minimum between access and backhaul transmission rates. 
Ref.~\cite{selim2018post}, instead, proposed a novel architecture relying on various types of drones, including a backhaul drone, as already discussed in Sec.~\ref{subsec:en_eff}.
In particular, the authors considered the backhaul drone to be connected via tether to an ad hoc truck operating as a satellite ground station, while a hybrid FSO/RF link connects the former to an untethered drone (specifically designed to provide access). \par
One of the main backhauling problems in post-disaster situations is to maximize the reliability of backhaul links, which requires perfect aligning between the transmit and receive antennas, given that mmWave systems are known for their high directivity.
In the context of emergency communications, splitting access and backhaul resources might be convenient since dynamic algorithms can be used to adapt the network depending on the different phases of disaster management; also, proper transmit power allocation combined with spatial multiplexing can mitigate interference from the user side.
However, accurate beam steering error models still need to be investigated and, since UAVs are also used as relays without any queuing capabilities, it also would be interesting to analyze spatio-temporal models with queuing for the ABSs~\cite{Kouzayha20hybrid}. \par


\subsection{Optimal Placement}
Given the increasing interest in vertical heterogeneous networks and their utility in post-disaster scenarios, it is essential to consider the optimal placement of ABSs because higher mobility and relocation flexibility characterize aerial nodes differently compared to their terrestrial counterparts. 
{ Moreover, it is evident that in emergency circumstances the problem of placement should not be constrained by any regulation (for example, on the ABSs' altitude) as long as it is meant to mitigate the damages.}
This, in turn, brings the opportunity to optimize the aerial nodes' locations to obtain the best network's performance either in terms of coverage, capacity, or energy efficiency (or any combination of the three). 
{ However, once again, the uncertainties in the network topology (\it{e.g.}, regarding the availability of access and backhaul links, the traffic distribution and intensity, \it{etc.}) can lead to very challenging optimization problems.} \par 
For instance,~\cite{herlich15guide} devised an analytical model describing how survivors can place (by means of a step-by-step guide in the form of mobile application) relay nodes such as mobile phones to communicate over long distances.
However, the main challenge is to improve the analytical model since it can only describe the trend of the channel quality and it is generally impractical to perform measurements in emergency scenarios.
Moreover, Ref.~\cite{Jowkar20quick} borrowed tools from optimization theory to determine how to place (\it{i.e.}, where and how many) edge-devices to conveniently provide reliable connectivity to FRs. \par
In the context of aerial networks, the choice of optimal placement of the ABS can be based on the study of several characteristics of the environment (\it{e.g.}, shape and size of the disaster area, QoR of the terrestrial network, load distribution, \it{etc.})~\cite{Matracia21disaster}. 
Knowing such information makes it possible to minimize the aggregate interference and maximize the QoS for both the FRs and the victims involved in the disaster.
Successful deployment of ABSs strongly depends on the assessment of the topology of the environment itself (which may take a relatively long time) and the estimation of the channel.
Hence, novel and efficient methods and algorithms need to be devised to solve these issues. \par
{ 
For general situations, authors in~\cite{ali20uav} introduced a setup with multiple TBSs supported by one single ABS, and optimized both the placement of the latter and the transmit power allocation of the nodes for both uplink and downlink operations.
Similarly,~\cite{kaleem22learning} proposed the \it{K-means and Q-learning assisted 3D ABS Placement and Power allocation algorithm (KQPP)} for sum-capacity maximization.
In particular the optimization framework used the K-means and discrete search algorithms for the ABS' ground coordinates, the model-free Q-learning reinforcement learning algorithm for power allocation (for varying ABSs' altitudes).
The convex optimization framework introduced in Ref.~\cite{valiulahi20multi} took into account also user scheduling and co-channel interference, showing promising results when compared to conventional methods.} \par
For emergency circumstances, the study presented in~\cite{Hayajneh18_cluster} focused on the impact of various parameters, such as the size of the recovery area, the altitude and the number of nodes per cluster, and the transmit power for backhaul on ABSs, while authors in~\cite{Zhao19} discussed the optimal hovering positions of UAVs-based relaying systems.
Moreover, works such as~\cite{Merwaday16} and~\cite{selim2018post} have quantitatively shown that how optimal placement of ABSs can contribute in terms of throughput enhancement. 
Finally,~\cite{Hydher20_intelligent} optimizes the positions of ABSs and the assigned users to maximize spectral efficiency while keeping a sufficient QoS for the users.
However, as the authors remarked, even assuming to solve the problem of positioning the UAVs, other challenges still need to be overcome, such as achieving sufficient self-organizing network capabilities, and mitigating the co-channel interference due to the high percentage of LoS aerial links.
Moreover, developing effective trajectory planning and control mechanisms is required in order to deal with the high mobility demands, as we will discuss in the following subsection.

\begin{table} []
    \centering
    \caption{Summary of the relevant papers on optimal placement}
    \begin{tabular}{|c|l|c|}
    \hline
    \textbf{Year} & \textbf{Main focus} & \textbf{Ref.} \\ \hline
    2015 & Long distance communications &~\cite{herlich15guide} \\\hline
    2016 & Throughput &~\cite{Merwaday16}\\ \hline
    \multirow{2}{*}{2018} & Stochastic-geometry-based performance analysis &~\cite{Hayajneh18_cluster} \\ \cline{2-3}
    & Coordination &~\cite{selim2018post} \\ \hline
    2019 & Relaying systems &~\cite{Zhao19} \\ \hline
    \multirow{4}{*}{2020} & Power allocation &~\cite{ali20uav} \\ \cline{2-3}
    & Scheduling, co-channel interference &~\cite{valiulahi20multi} \\ \cline{2-3}
    & Edge-devices for FRs &~\cite{Jowkar20quick} \\ \cline{2-3}
    & Spectral efficiency &~\cite{Hydher20_intelligent} \\ \hline 
    2021 & Ergodic capacity &~\cite{Matracia21disaster} \\ \hline
    2022 & Sum capacity &~\cite{kaleem22learning} \\ \hline
    \end{tabular}
    \label{tab:placement}
\end{table}

\subsection{Optimal Trajectory} \label{subsec:traj}
The research on optimizing ABSs' trajectory for post-disaster networks is still in its infancy and needs further investigation.
Trajectory planning is critical when deploying ABSs, especially those characterized by fixed wings because they cannot simply stay aloft in a specified position.
Therefore, the location of the aerial nodes should be adjusted in real-time, in order to avoid collisions while following the movements of the load (\it{e.g.}, users escaping from a building on fire and moving towards a shelter).
One possible direction in this domain consists in developing powerful distributed algorithms that can identify optimal trajectories for a fleet of UAVs (which can be needed in case of large disasters or high capacity needs) and a team of FRs or a group of victims that need to be rescued.
Further considerations on UAV trajectory optimization are available in~\cite{Naqvi18key}. \par
Authors in~\cite{wang20safe} recently studied the influence of obstacles on the status of road networks and the speed of rescue vehicles to generate effective safe routes for FRs. 
In this work, the authors mentioned several challenges to be overcome in the future, for example: \it{(i)} extending the framework to the case with multiple vehicles and destinations; \it{(ii)} developing a routing algorithm that takes into account not only the risks associated to routes, but also FRs' time constraints; \it{(iii)} developing a user interface that allows to adapt the routing parameters based on real situations.
Instead,~\cite{arafat19localization} presented swarm-intelligence-based localization and clustering algorithms for trajectory optimization in aerial emergency networks. 
According to the authors, the next step would be to investigate bio-inspired AI techniques to improve the performances of the algorithms. 
Another interesting work, presented in Ref.~\cite{christy17path}, aims to optimize the UAV flying paths, in terms of both coverage area and energy efficiency, depending on the type of disaster (\it{e.g.}, earthquake, flood, wildfire, \it{etc.}).
In addition, an extensive comparison between the \it{S-}, \it{O-}, \it{ZigZag-}, and rectangular paths was discussed by the authors. 

\subsection{Scheduling}
{ Another challenging aspect of PDCs is transmission scheduling, due to multiple issues. 
For example, it would be impossible to perform opportunistic scheduling if the transmitter is not able to continuously obtain the channel state information (CSI); evidently, this problem would occur frequently given the current difficulties in identifying the proper channel model (see Sec.~\ref{subsec:channelMod}) and the actual network topology.
Another open challenge consists in determining priorities depending on the type of user (for instance, the ideal scheduling algorithm should work so that trapped victims and first responders are better served than other types of users).
}
Works such as~\cite{Zhao19} jointly optimized UAVs' trajectory and scheduling to connect ground devices and surviving BSs in the most effective way.
On the other hand, authors in~\cite{arivudainambi14heuristic} devised a domain-specific memetic algorithm to solve broadcast scheduling problem in wireless mesh networks.
The effectiveness and efficiency of this algorithm were then validated via simulations.
Finally,~\cite{gao21dynamic} introduced a dynamic priority scheduling strategy for UAV-assisted MANETs.
In particular, the Gauss-Markov mobility model captures the nodes' dynamics while the devised priority assignment takes into account both the delay experienced by the packet and the effects of such delay on the next transmissions. 

\subsection{HO Management}
Another major challenge in PDCs is the HO management between the ABSs, especially in terms of HO cost (\it{i.e.}, the fraction of time that, on average, is spent for associating with another ABS).
Note that for the case of aerial networks, this does not depend only on the user's velocity but also on the velocity of the aerial nodes since, as extensively discussed in the previous subsection, they are usually non-stationary.
{ In addition, the emergency situation may easily lead to an increased HO rate, since the users may be moving as fast as they can to save their lives and the increased traffic demand may require considerable network densification.
This, in turn, would imply a decrease in terms of QoS and energy efficiency.
} \par
Authors in~\cite{ray15LTE} devised an LTE-advanced (LTE-A) user-equipment-controlled
and BS-assisted handover scheme to prolong the network's autonomy.
Ref.~\cite{arshad2018integrating} investigated the rate performance of a UAV-aided three-tier downlink network by taking into account the effect of HO rates on the user rate. 
In particular,~\cite{arshad2018integrating} assumed two operating modes, namely conventional (where all the BSs transmit data and control signals) and control/data split (where only UAVs provide control signals but any BS can provide the data), showing that the latter generally reduces the HO cost.
To this extent, designing efficient HO schemes for integrated terrestrial-vertical heterogeneous networks becomes very challenging.
For a given situation, the main problem is, indeed, to understand the optimal density of nodes so that the HO rate remains acceptable.

\subsection{Content Caching}
Content caching is an emerging area for data distribution where the content is provided to the user from the closest servers~\cite{bernardini16caching}.
In general, the potential benefits provided by edge caching consist in: \it{(i)} caching popular contents to reduce the delay and load experienced by the BSs (assuming that the edge nodes are placed close enough to the users), and \it{(ii)} minimizing traffic through backhaul links (due to redundant data transmissions avoidance) \cite{su18secure}.
{ On the downside, if data are gathered from various sources (for example, sensors or social networks), then it is necessary to bridge any discrepancy or error before sharing.
Modern technologies can already deal with this issue in ordinary occasions, but devising a similar framework for PDCs is much more challenging~\cite[Sec. 7]{khan22emerging}.}
We recall that a recent review on fog computing can be found in~\cite{Kurniawan21fog}.
In emergency scenarios, we strongly believe that content caching can be an important tool to timely share vital information such as safe routes, shelters, and specific safety guidelines. \par
Authors in~\cite{su18secure} proposed an effective and secure caching scheme for ensuring backup in case of disasters.
This scheme was designed to serve mobile users in mobile social networks (MSNs) with fog computing (\it{i.e.}, one of the main aspects of edge computing) and secure encryption. 
Instead, authors in~\cite{zhou20edge} used edge caching, where a generic multiple UAV-enabled RAN (UAV-RAN) is utilized for spectral efficiency enhancement.
In particular, they assumed a stochastic geometry-based setup and suggested dividing the caching contents into two sets with different ranges of popularity: the most popular files would be stored at all ABSs, while the least ones just at one ABS.
The optimization problem consists in finding the proper popularity threshold to define the two sets, and the authors derived the expression of the threshold based on Fenchel duality.
One major challenge for using content caching in post-disaster situations is optimizing content distribution in cloud servers to reduce costs, latency, and resource consumption. 

\subsection{IRS-Enabled Networks}
{ Two important challenges in emergency scenarios arise from the rich scattering (see Sec.~\ref{subsec:channelMod}) environment as well as the scarcity of surviving infrastructure.
IRSs could play an important role in overcoming these challenges, especially when combining their co-located and separated operation mode.
In fact, although the IRS is usually designed as a whole device, it is possible to engineer detachable patches that, if necessary, can operate independently or even collaborate (if they are not too far from each other).
In this way, these elements can mitigate the effects of both scattering and multipath propagation.
Nonetheless, the inherent and tight constraints in terms of memory, computational capability, and energy may prevent from applying this technology~\cite{kisseleff21reconfigurable}.
Another interesting application of IRSs derives from their installation on UAVs, as suggested in~\cite{alfattani21aerial}.
The main related challenges for this technology derive from \it{(i)} the need of continuous reconfiguration of the IRS in order to perform accurate channel estimation while the UAV is moving and \it{(ii)} the limited payload capability of untethered UAVs.}\par
Several contributions in this topic were proposed by the authors of Ref.~\cite{kisseleff21reconfigurable}, where IRS technology was reviewed, discussed, and evaluated for various challenging environments (\it{i.e.}, post-disaster, industrial, underwater, and underground).
Moreover, authors in~\cite{jaafar22enhancing} discussed IRS-assisted and IRS-equipped UAV PSNs, for which aerial surveillance and SAR applications were respectively proposed. \par
For more general applications, the study presented in~\cite{cai20resource} focused on the use of an IRS-assisted UAV for serving multiple ground users. 
The authors jointly optimized the UAV's trajectory and velocity, resource allocation strategy, and phase control at the IRS in order to minimize the overall power consumption\footnote{The latter work is general and not limited to post-disaster scenarios.}.
We also recall \cite{yao21resource} in Sec.~\ref{subsubsec:chann_LAPs} for the problem of characterizing the link between UAV-mounted IRSs and trapped users.
The imperfection of CSI can be captured by the framework proposed in~\cite{hu21robust}, which focused on robust and secure IRS-assisted multi-user multiple-input single-output (MISO) downlink communications.
The authors assumed self-sustainable IRSs (capable to simultaneously reflect and harvest energy from the received signals) and jointly designed beamformers, IRS phase shifts, and energy harvesting schedule in order to maximize the system sum-rate.

\section{Conclusions} \label{sec:conclusion}
In this paper, we reviewed recent works on the development of post-disaster wireless communication networks, which are envisioned as one of the most relevant aspects of 5G and beyond connectivity.
We showed that current research is highly focused on developing novel algorithms, technologies, and architectures to improve the overall network performances in case of emergency. Moreover, we discussed the main challenges and limitations these studies faced, and linked them to future research directions, as well as the need to carry on practical experiments and field tests.
Finally, by numerical evaluation of the coverage probability in realistic post-disaster scenarios, we proved the effectiveness of creating novel ad hoc architectures that properly combine existing wireless technologies. \par
Therefore, by discussing relevant literature works, novel case studies, and current technological challenges, we hope this work provides a valuable bird's eye view for both academic and industrial researchers who wants to explore the area of PDCs.

\bibliographystyle{IEEEtran}
\bibliography{bibliography.bib}

 \begin{IEEEbiography}
 [{\includegraphics[width=0.96in,height=1.25in,clip]{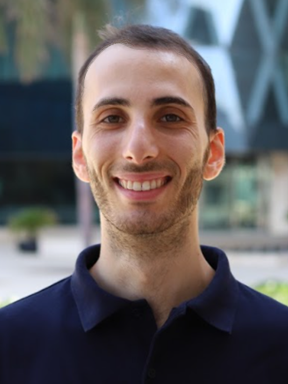}}]
 {Maurilio Matracia}
 [S'21] is a Doctoral Student at the Communication Theory Lab (CTL), King Abdullah University of Science and Technology (KAUST), Kingdom of Saudi Arabia (KSA). 
 He received his M.Sc. degree in Electrical Engineering from the University of Palermo (UNIPA), Italy, in 2019.
His main research interest regards rural and emergency communications.
\end{IEEEbiography}


\begin{IEEEbiography}
[{\includegraphics[width=0.96in,height=1.25in,clip]{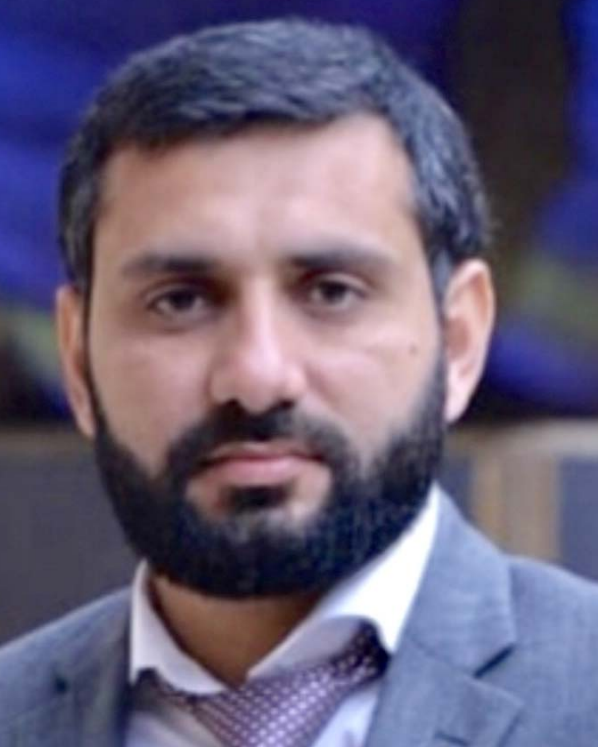}}]
{Nasir Saeed}
[S'13 M'15 SM'19] received the Ph.D. degree in Electronics and Communication Engineering from Hanyang University, Seoul, South Korea, in 2015.
He is currently an Associate Professor with the Department of Electrical Engineering, Northern Border University, Arar, KSA. His current research interests include cognitive radio networks, underwater wireless communications, aerial networks, dimensionality reduction, and localization.
\end{IEEEbiography}


 \begin{IEEEbiography}
 [{\includegraphics[width=0.96in,height=1.25in,clip]{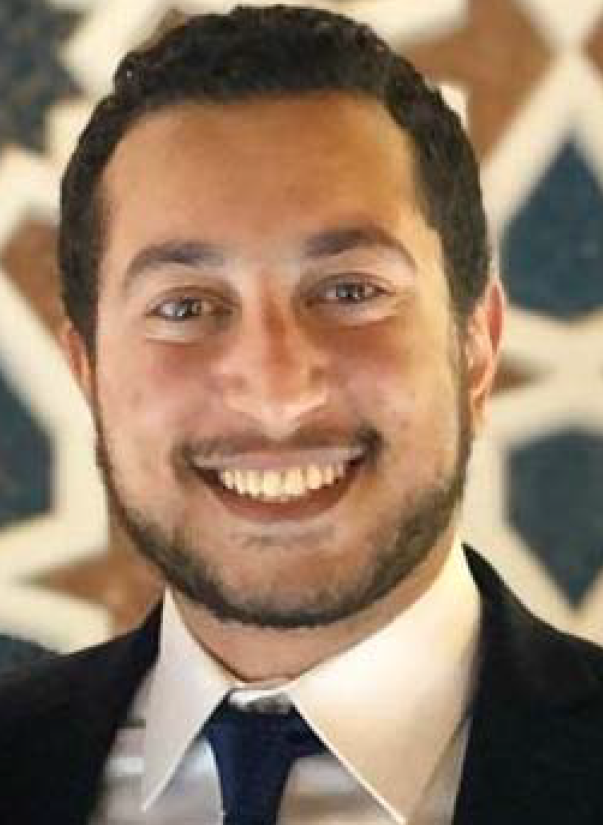}}]
 {Mustafa A. Kishk}
[S'16, M'18] is currently an Assistant Professor at the Electronic Engineering Department, Maynooth University, Ireland. 
 He received his B.Sc. and M.Sc. degree from Cairo University in 2013 and 2015, respectively, and his Ph.D. degree from Virginia Tech in 2018.
 His research interests include stochastic geometry, energy harvesting wireless networks, UAV-enabled communication systems, and satellite communications.
 \end{IEEEbiography}


 \begin{IEEEbiography}
 [{\includegraphics[width=0.96in,height=1.25in,clip]{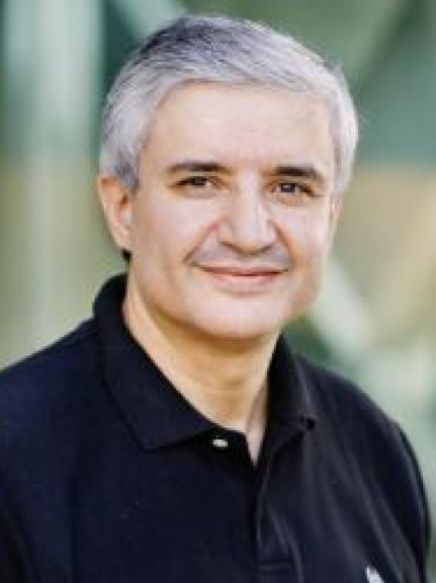}}]
 {Mohamed-Slim Alouini} 
 [S'94, M'98, SM'03, F'09] was born in Tunis, Tunisia.
 He received the Ph.D. degree in Electrical Engineering from the California Institute of Technology (Caltech), Pasadena, USA, in 1998. He served as a faculty member in the University of Minnesota, Minneapolis, USA, and in the Texas A$\&$M University at Qatar, Doha, Qatar, before joining KAUST as a Professor of Electrical Engineering in 2009. 
 His current research interests include the modeling, design, and performance analysis of wireless communication systems.
 \end{IEEEbiography}

\end{document}